\documentclass[a4paper,fleqn,usenatbib,useAMS]{mnras}

\usepackage{graphicx}	
\usepackage{amsmath}	
\usepackage{amssymb}	
\usepackage{multicol}        
\usepackage{bm}		
\usepackage{pdflscape}	

\usepackage{enumerate}
\usepackage{mathtools}

\usepackage[shortlabels]{enumitem}

\defcitealias{McClure-Griffiths_2010}{McG10}

\newcommand{\lsim}{{\;\raise0.3ex\hbox{$<$\kern-0.75em\raise-1.1ex\hbox{$\sim$}}\;}}
\newcommand{\gsim}{{\;\raise0.3ex\hbox{$>$\kern-0.75em\raise-1.1ex\hbox{$\sim$}}\;}}

\usepackage[T1]{fontenc}
\usepackage{ae,aecompl}

\usepackage{newtxtext,newtxmath}

\title[Clumpy clouds in magnetized halo]{Magnetic field draping around clumpy high-velocity clouds in galactic halo}

\author[Jung et al.]{S. Lyla Jung$^{1}$\thanks{e-mail: \href{mailto:lyla.jung@anu.edu.au}{lyla.jung@anu.edu.au}}, Asger Gr{\o}nnow$^{2}$, N. M. McClure-Griffiths$^{1}$
\\
\\
$^{1}$ Research School of Astronomy \& Astrophysics, The Australian National University, Canberra ACT 2611, Australia
\\
$^{2}$ Kapteyn Astronomical Institute, University of Groningen, Landleven 12, 9747 AD Groningen, The Netherlands}

\date{Last updated}

\pubyear{2022}

\begin{document}
\label{firstpage}
\pagerange{\pageref{firstpage}--\pageref{lastpage}}
\maketitle

\begin{abstract}
Throughout the passage within the Galactic halo, high-velocity clouds (HVCs) sweep up ambient magnetic fields and form stretched and draped configurations of magnetic fields around them. Many earlier numerical studies adopt spherically symmetric uniform-density clouds as initial conditions for simplicity. However, observations demonstrate that HVCs are clumpy and turbulent. In this paper, we perform 3D magnetohydrodynamic simulations to study the evolution of clouds with initial density distributions described by power-law spatial power spectra. We systematically study the role of (i) the initial density structure, (ii) halo magnetic fields, and (iii) radiative cooling efficiency upon infalling HVCs. We find that (i) the clouds' density structure regulates mixing and mass growth. Uniform clouds grow from the onset of the simulations while clumpy clouds initially lose gas and then grow at later times. Along the same lines, the growth curve of clumpy clouds depends on the slope of the initial density power spectra. (ii) Magnetic fields suppress hydrodynamic instabilities and the growth of small-scale structures. As a result, magnetized clouds develop long filaments extended along the streaming direction whereas non-magnetized clouds are fragmented into many small clumps. (iii) Efficient cooling keeps the main cloud body more compact and produces decelerated dense clumps condensed from the halo gas. This work potentially helps us understand and predict the observed properties of HVCs such as the detectability of magnetized clouds, the presence of decelerated HI structures associated with HVC complexes and small-scale features, and a possible link between the origin and the fate of HVCs.
\end{abstract}





\section{Introduction}

High-velocity clouds (HVCs) are part of the multi-phase circumgalactic medium (CGM) of the Milky Way. Though it is the hot gas ($T\sim10^{6}\,\rm K$; \citealt{Henley_2015}) that fills much of the volume of the Galactic halo, a significant portion of the CGM is in cooler gas phases, e.g., neutral hydrogen (HI). 
The observed sky coverage of HI HVCs is $\sim 15\%$ with the column density threshold of $N_{\rm HI}>2\times10^{18}\,\rm cm^{-2}$ (\citealt{Westmeier_2018}) and this increases to $\sim 37\%$ with a lower threshold $N_{\rm HI}>7\times10^{17}\,\rm cm^{-2}$ (\citealt{Murphy_1995}; \citealt{Lockman_2002}).
The total mass estimate of the observed HVCs is largely dependent on the accuracy of the distance measurements to the clouds.
\citet{Putman_2012} estimate the total mass of 
HVCs not associated with the Magellanic System to be $\sim 7.4\times10^{7}\,\rm M_{\odot}$. When the Magellanic system (the Magellanic Stream, Leading Arm, and Bridge) is included, the estimated HI mass is an order of magnitude higher.

These cold-phase gas clouds are a potential source of gas that could fuel the future star formation activity of the Galaxy.
The gas inflow and outflow rates in the Milky Way halo are often estimated from the observed Galactic Standard of Rest velocities of HVCs (inflow: $v_{\rm GSR} < 0\,\rm km s^{-1}$, outflow: $v_{\rm GSR} > 0\,\rm km s^{-1}$). 
Considering HI HVCs including the Magellanic Stream, \citet{Richter_2012} gives the inflow rate $\dot{M}_{\rm in}\gsim0.7\,\rm M_{\odot}yr^{-1}$.
Similarly, \citet{Fox_2019} estimate the inflow rate $\dot{M}_{\rm in}\gsim0.53\,\rm M_{\odot}yr^{-1}$ and the outflow rate $\dot{M}_{\rm out}\gsim0.16\,\rm M_{\odot}yr^{-1}$ excluding the contributions of the Magellanic Stream and the Fermi Bubbles. Similar results are also presented in \citet{Clark_2021}.
It is clear that the Milky Way not only has a very apparent accretion of the Magellanic system, but also the non-Magellanic origin HVCs are overall inflowing into the Galaxy rather than outflowing. Such inflow might be the dominant source of gas needed to sustain the star formation activity of the Milky Way.

However, it is still an open question how cold clouds can survive the dynamic environment of the CGM and deliver gas to the Galaxy. 
Surrounded by hot halo gas, clouds falling towards the Galaxy are under severe ram pressure that can strip gas away from the cloud. Also, hydrodynamic instabilities (typically, Rayleigh-Taylor and Kelvin-Helmholtz instabilities) provoke mixing across the cloud's interface between the two media with different densities and velocities (e.g., \citealt{Fielding_2020}).
Observed head-tail morphology (\citealt{For_2013}) and complex small-scale structures arising across the edges of HVCs (\citealt{Barger_2020}) are direct evidence of ongoing ablation of the clouds moving at high velocities.

On the other hand, the mixing of cold and hot gas produces gas at intermediate density and temperature where radiative cooling can be efficient. 
Under certain conditions where the cooling is faster than the mixing and stripping, the total cold gas mass of the cloud system can grow with time (e.g., \citealt{Marinacci_2010}; \citealt{Armillotta_2016}; \citealt{Gronke_2018}; \citealt{Li_2020}; \citealt{Kanjilal_2021}; \citealt{Heitsch_2021}; \citealt{Gronnow_2022}).

Introducing further complications, magnetic fields in the halo significantly affect the overall evolution of the clouds.
When a cloud is moving super-Alfvenically, it sweeps up the magnetic field lines in the halo and develops a strongly magnetized layer at the cloud-halo interface.
The effect of magnetic fields on the survivability of clouds moving at high velocities has been actively studied from the numerical simulations' perspective
(\citealt{Jones_1996}; \citealt{Gregori_1999, Gregori_2000}; \citealt{Santillan_1999}; \citealt{Dursi_2008}; \citealt{Shin_2008}; \citealt{Kwak_2009}; \citealt{McCourt_2015}; \citealt{Banda-Barragan_2016}; \citealt{Gronnow_2017}; \citealt{Banda-Barragan_2018}; \citealt{Cottle_2020}; \citealt{Sparre_2020}; \citealt{Gronnow_2022}). Earlier models clearly show that the magnetized layer provides tension and suppresses hydrodynamic instabilities across the field lines.
Yet, the extent of stability provided by magnetic fields to the clouds is not well constrained as it depends on, for example, types of instabilities, the strength of the magnetic field, and the direction of the field with respect to the clouds' motion. 
Earlier work by \citet{Gregori_1999} find enhanced Rayleigh-Taylor instability along the direction perpendicular to the cloud's motion and the initial ambient magnetic field.
More recent works with higher-resolution simulations give mixed results especially when radiative cooling is included. For example, \citet{Sparre_2020} and \citet{Cottle_2020} find only a small effect of magnetic fields on the survival of the clouds while \citet{McCourt_2015} and \citet{Gronnow_2018} predict a large effect. In either case, studies agree that the magnetic field does not lead to faster destruction.

Many of the studies mentioned above, with notable exceptions of \citet{Banda-Barragan_2018,Banda-Barragan_2019}, assume a simple spherical cloud with a uniform density core for initial conditions.
In such a setting, the size of the cloud is one of the important deciding factors of the survival timescale of the clouds. 
For example, \citet{Li_2020} present that the lifetime of their simulated clouds closely follows a simple power-law dependency on the cloud size and other properties (see their equation 19 for the exact form of the fitting function).
However, observations clearly show that HVCs are in fact clumpy and turbulent, similar to much of the interstellar medium (ISM); a large cloud complex may have multiple dense cores and the spatial power spectrum of the column density distribution follows a power law distribution (e.g., \citealt{Marchal_2021}).
Hence, there is a need to examine to what extent predictions from spherical cloud simulations are valid when it comes to clouds with more realistic structures.
There have been a handful of simulations probing the evolution of clouds with complex and realistic density structures, however, many of them do not include magnetic fields (e.g., \citealt{Patnaude_2005}; \citealt{Cooper_2009}; \citealt{Schneider_2017}) or are not in the parameter regime (for example, in terms of metallicity or relative velocity) applicable to clouds falling into the halo, like HVCs (e.g., \citealt{Banda-Barragan_2018,Banda-Barragan_2019}).

In this paper, we investigate the properties and the evolution of clouds that have an initial density distribution motivated from observations.
We choose the physical parameters of the fiducial simulation, e.g., gas metallicity, cloud velocity, and density structure, to be in line with the observed HVCs. 
More details on simulation settings are in Section \ref{sec:method}.
In Section \ref{sec:survivability}, we compare the evolution of the total cold gas mass of each model and evaluate the survival of the clouds under various physical conditions.
In Section \ref{sec:3}, we discuss processes that shape the evolution of the clouds by comparing the structural properties of the clouds in different models.
Discussions on the resolution effect, caveats, and limitations in interpreting our models are in Section \ref{sec:resol} and Section \ref{sec:caveat}.
We connect the results of the simulations to the observed properties of the Milky Way HVCs and make predictions for future observations in Section \ref{sec:milky_way}.
Section \ref{sec:summary} is the summary and conclusion.

\section{Methodology}\label{sec:method}
\subsection{Simulations overview}\label{sec:simul_overview}

\begin{figure*}
    \centering
    \includegraphics[width=0.9\textwidth]{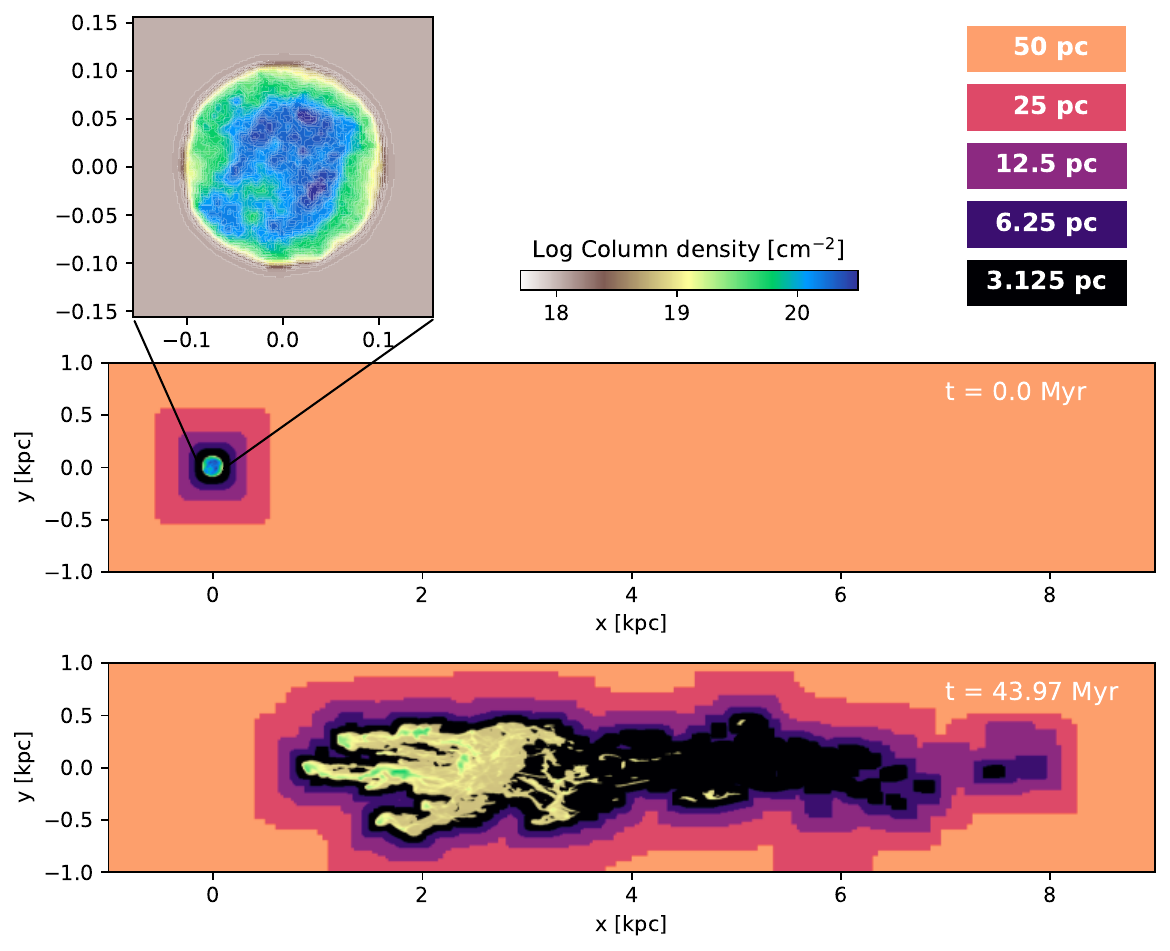}
    \caption{
    The zoom-in panel on the top shows the column density distribution of the initial clumpy cloud. 
    The bottom two panels show the distribution of the column density and the AMR refinement levels of the MC model (see Table \ref{tab:t1}) at the initial and the final snapshot ($t = 0$ and $44\,\rm Myr$). 
    In both panels, the column density distributions share the same colour scale as the top panel and only the gas with $C>0.25$ is presented, where $C$ is a passive scalar tracer that follows the advection of the original cloud material (see equation \ref{eq:tracer} and the related text for the definition of $C$).
    We confirm that any structures at $C>0.25$ are covered by AMR cells at the maximum refinement level (black region) throughout the evolution of the cloud.
    }  
    \label{fig:init}
\end{figure*}

\begin{table*}
\caption{A list of simulations and input parameters. Column 1 is the name of the models, column 2 is the type of cloud used as the initial condition, and columns 3 to 6 are the gas velocity, density, metallicity, and magnetic field strength, respectively. Column 7 is the length of the smallest AMR cell in the simulation domain, column 8 shows whether radiative cooling is implemented, and column 9 is the slope of the initial SPS of the clouds.
}\label{tab:t1}
\begin{minipage}{\textwidth}\centering
\begin{tabular}{ccccccccc}
& Cloud type & v$_{\rm halo}$ ($\rm km s^{-1}$) & $\rho_{\rm halo}$ ($\rm g cm^{-3}$) & Metallicity (Z$_{\odot}$) & B$_{\rm 0}$ ($\rm \mu G$) & Max resolution (pc)& Cooling & Slope\\\hline \hline
MC & Clumpy-R32 & 200 & 10$^{-27}$ & 0.3 & 0.1\footnote{This corresponds to plasma $\beta=P_{\rm thermal}/P_{\rm B}\approx350$.} & 3.125 & Yes & -3\\ \hline 
HC & Clumpy-R32 & 200 & 10$^{-27}$ & 0.3 & 0 & 3.125 & Yes& -3\\ \hline 
MU & Uniform & 200 & 10$^{-27}$ & 0.3 & 0.1 & 3.125 & Yes & -\\ \hline 
HU & Uniform & 200 & 10$^{-27}$ & 0.3 & 0 & 3.125 & Yes & -\\ \hline 
MC-Z003& Clumpy-R32 & 200 & 10$^{-27}$ & 0.03 & 0.1 & 3.125 & Yes & -3\\ \hline 
MC-NoCool & Clumpy-R32 & 200 & 10$^{-27}$ & - & 0.1 & 3.125 & No & -3\\ \hline 
MC-p2.7 & Clumpy-p2.7 & 200 & 10$^{-27}$ & 0.3 & 0.1 & 3.125 & Yes & -2.7\\ \hline 
MC-p3.5 & Clumpy-p3.5 & 200 & 10$^{-27}$ & 0.3 & 0.1 & 3.125 & Yes & -3.5\\ \hline 
MC-R16 & Clumpy-R16 & 200 & 10$^{-27}$ & 0.3 & 0.1 & 6.25 & Yes & -3\\ \hline
MC-R64 & Clumpy-R64 & 200 & 10$^{-27}$ & 0.3 & 0.1 & 1.5625 & Yes & -3\\ \hline
\end{tabular}
\end{minipage}
\end{table*}

\begin{table*}
\begin{minipage}{0.7\textwidth}\centering
\caption{A list of initial cloud types used in this study. Column 1 is the name of the clouds, column 2 is $\chi_{\rm mean} \equiv \langle{\rho}_{\rm cloud}\rangle/\rho_{\rm halo}$, where $\langle{\rho}_{\rm cloud}\rangle$ is the mean density of all gas cells with $C=1$. See equation \ref{eq:tracer} and the related paragraph for the definition of $C$. Column 3 is $\chi_{\rm median} \equiv \tilde{\rho}_{\rm cloud}/\rho_{\rm halo}$, where $\tilde{\rho}_{\rm cloud}$ is the median density of all gas cells with $C=1$. Column 4 is the cold gas mass of the clouds and column 5 is the total gas mass within $0.1\,\rm kpc$ from the cloud centre where $C$ is set to 1.
 }\label{tab:init}
\begin{tabular}{ccccc}
Cloud type & $\chi_{\rm mean}$ & $\chi_{\rm median}$ & 
\begin{tabular}[c]{@{}c@{}} $M_{\rm cloud}$ [}$\rm M_{\odot}${]\\  {($T<2\times10^{4}\,\rm K$)}{}\end{tabular}
& \begin{tabular}[c]{@{}c@{}} $M_{\rm cloud}$ [}$\rm M_{\odot}${]\\  {($C=1$)}{}\end{tabular}\\ \hline\hline
Uniform    & 181 & 197 & $1.21\times10^{4}$ & $1.12\times10^{4}$ \\ \hline
Clumpy-R16 & 198 & 149 & $1.23\times10^{4}$ & $1.22\times10^{4}$  \\ \hline
Clumpy-R32 & 200 & 142 & $1.23\times10^{4}$ & $1.23\times10^{4}$ \\ \hline
Clumpy-R64 & 200 & 133 & $1.23\times10^{4}$ & $1.23\times10^{4}$  \\ \hline
Clumpy-p2.7 & 199 & 104 & $1.23\times10^{4}$ & $1.21\times10^{4}$\\ \hline
Clumpy-p3.5 & 199 & 181 & $1.23\times10^{4}$& $1.24\times10^{4}$\\ \hline
\end{tabular}
\end{minipage}
\end{table*}

We perform a set of 3D magnetohydrodynamic (MHD) simulations of moving clouds using the {\sc PLUTO} code version 4.1 (\citealt{Mignone_2012}). Assuming infinite conductivity, the fluid and magnetic fields in the simulation box evolve following the ideal MHD equations. We use the HLLC Riemann solver (\citealt{Toro_1994}) which has a good balance of accuracy and stability for all our simulations.
We have run our fiducial MHD simulation using HLLD solver as well because the HLLD solver is often considered to be a more accurate extension of HLLC for MHD cases. We have verified that the differences between the models with HLLC and HLLD solvers are not significant in our simulation setting\footnote{Later in Fig. \ref{fig:time}, we show the cold gas evolution in the HLLD run.}.
We apply the hyperbolic divergence-cleaning algorithm introduced by \citet[see also \citealt{Mignone_2010}; \citealt{Mignone_2010(2)}]{Dedner_2002}, that approximately enforces the solenoidal constraint of magnetic fields ($\nabla \cdot \textbf{B} = 0$).
 
The simulation domain is a rectangular box elongated along the $x$-axis, i.e., ($L_{\rm x}, L_{\rm y}, L_{\rm z}) = (10, 2, 2)\,\rm kpc$. Boundary conditions are set to outflow, except at the gas injection boundary at $x=-1\,\rm kpc$. From the injection boundary, hot halo gas is injected along the $x$-axis at a constant positive velocity ($v_{\rm halo}$).
Initially, we place an overdense structure (the ``cloud") centred at $1\,\rm kpc$ away from the injection boundary (i.e., $x=0\,\rm kpc$). See the middle panel of Fig. \ref{fig:init} for the initial setting of the simulation domain.
The rest of the volume is filled with halo gas 
with the same properties as the medium injected at the boundary.
The cloud is initially static with respect to the simulation domain and in pressure equilibrium with the halo gas.

We include radiative cooling via collisional ionization equilibrium based on the tables of \citet{Sutherland_1993}, except for a model in that we turned off the cooling for comparison study. 
The cooling rate of gas depends on the metallicity and temperature of the gas. Cooling is switched off below a temperature threshold of $10^{4}\,\rm K$.
This artificial cooling floor is adopted to indirectly consider the gas heating effect caused by the UV background.
The metallicity is uniform throughout the simulation domain.

We use the adaptive mesh refinement (AMR) technique to ensure that any dense structures in the simulations are covered with high-resolution grids throughout the time domain of the simulations.
For the standard resolution, we use a base grid of (200, 40, 40) cells along each axis and then allow refinement by up to 4 levels based on the density gradient.
Each level refines cells to half the cell length of the previous level. Thus, the minimum AMR cell size is
$(10 \,\rm{kpc}/200) \times 2^{-4} = 3.125\,\rm pc$, equivalent to 32 cells per cloud radius.
See Fig. \ref{fig:init} for the distribution of cells in different refinement levels at the initial (upper panel) and the final (lower panel) snapshot of the fiducial simulation run.

Table \ref{tab:t1} summarises the input parameters of the models in this study.
In all models, the density of the halo gas is $\rho_{\rm halo}=10^{-27}\,\rm g cm^{-3}$, i.e., the particle number density $\sim 0.001\,\rm cm^{-3}$ and the temperature $8.5\times10^{5}\,\rm K$ that are representative of the inner Milky Way corona (e.g., \citealt{Henley_2015}; \citealt{Miller_2015}). The initial velocity of the clouds with respect to the surrounding halo gas is $200 \,\rm km s^{-1}$, which corresponds to a Mach number
\begin{equation}
    \mathcal{M_{\rm halo}}=\frac{v_{\rm halo}}{c_{\rm s}}=1.3,
\end{equation}
where the sound speed of the halo gas is $c_{\rm s}=\sqrt{\frac{5P}{3\rho_{\rm halo}}}$ as we consider all gas in our models as a monatomic ideal gas and $P$ is the gas pressure.
This choice of velocity is motivated by the deviation velocity, i.e., the deviation of observed clouds' velocity from the Galactic rotation velocity, of observed HVCs (\citealt{Moss_2013}).

The initial simulation domain has uniform magnetic field strength $B_{\rm 0}=0.1\,\rm\mu G$, unless specified as non-magnetized models.
This choice of the magnetic field strength represents the halo gas $10\,\rm kpc$\footnote{This is a typical distance to observed Milky Way HVCs (e.g., \citealt{Wakker_2001}; \citealt{Thom_2008}).} above the Galactic disk in the solar neighbourhood (i.e., $8\,\rm kpc$ radius from the Galactic centre) according to the Galactic magnetic field model of \citet{Jansson_2012}.
Physically, this corresponds to the plasma $\beta=P_{\rm thermal}/P_{\rm B}\approx350$ and places the medium in the super-Alfvenic regime\footnote{The Alfvenic Mach number $\mathcal{M_{\rm A}} = v_{\rm halo}/v_{\rm A}\approx 22.4$, where the Alfven velocity is $v_{\rm A} = \sqrt{\frac{B_{\rm 0}^{2}}{4\upi\rho_{\rm halo}}}\approx8.9\,\rm km s^{-1}$}. 
The direction of the halo magnetic field lines is perpendicular to the direction of the cloud motion.
Earlier studies by \citet{Banda-Barragan_2016} and \citet{Gronnow_2018} show that only the transverse component drapes around the cloud and is amplified even when there are magnetic field components both parallel and transverse to the motion. 
As a result, the transverse component tends to dominate the evolution for all field orientations in most cases, other than when the halo magnetic fields are very close to parallel with negligible transverse magnetic fields.\\

\setlength{\parindent}{0cm}
\large{\underline{\textbf{Models}}}
\begin{itemize}[topsep=0pt]
    \item The MC model is a fiducial simulation representing typical observed HVCs in the Milky Way halo. The metallicity is set to $0.3\,\rm Z_{\odot}$ which is motivated by the typical range of metallicities of observed HVCs (e.g., Complex A $\sim0.1\,\rm Z_{\odot}$, Complex C $0.1-0.3\,\rm Z_{\odot}$, Magellanic Stream $0.1-0.3\,\rm Z_{\odot}$, Smith cloud $\sim0.5\,\rm Z_{\odot}$; from reviews by \citealt{Putman_2012} and \citealt{Richter_2017}, and references therein). The cloud in this model has a clumpy density structure. Details on how we generate the initially clumpy cloud will be presented shortly in Section \ref{sec:clumpy_setting}.
    
    \item The HC model is a pure hydrodynamic (HD) counterpart of the fiducial model. All the physical parameters are identical to the MC model except that there are no magnetic fields. See Section \ref{sec:mhd_vs_hd} for the detailed comparison of the evolution of clouds with and without magnetic fields.
    
    \item The MU model and the HU model are MHD and HD simulations with a spherically symmetric initial density cloud instead of a clumpy cloud.
    
    \item For testing the effect of radiative cooling on the evolution of clouds, we have the MC-Z003 model with low metallicity ($0.03\,\rm Z_{\odot}$, i.e., less efficient cooling) throughout the simulation volume. This metallicity roughly corresponds to the metallicity of HVCs with pristine intergalactic origin. The MC-NoCool model has radiative cooling turned off completely. 
    
    \item We study the effect of varying the slope of the initial density spatial power spectrum (SPS) by comparing the MC-p2.7 model and the MC-p3.5 model which have a spectral slope shallower ($-2.7$) and steeper ($-3.5$) than the MC model ($-3$), respectively. The choice of slopes used in this study is made within the observed range of the slopes in the Galactic ISM presented in \citet{Pingel_2018} and references therein. See Section \ref{sec:clumpy_setting} on how the structure of a clumpy cloud changes with the varying spectral slope.
    
    \item We test the resolution effect using models with a maximum resolution lower (MC-R16; 16 cells per cloud radius) and higher (MC-R64; 64 cells per cloud radius) than the fiducial MC model (i.e., 32 cells per cloud radius).
\end{itemize}

\subsection{Initial cloud density distribution} \label{sec:clumpy_setting}

For our ``clumpy'' cloud simulations, we initialise the density structure of the cloud to resemble the observed log-normal density distribution of gas clouds.
As a reference, we use a $21\,\rm cm$ HI map of the Smith Cloud observed with the Robert C. Byrd Green Bank Telescope of the National Radio Astronomy Observatory from \citet{Lockman_2008}.
We parameterise the properties of the cloud based on the column density SPS and probability distribution function (PDF).
Only pixels with a column density above 26 times the noise level (i.e., $N_{\rm HI} >10^{19.8}\,\rm cm^{-2}$) are considered in this process. 
Like many other areas of the observed ISM (e.g., \citealt{Pingel_2018}), the SPS of the HVC can be readily approximated in a power-law form $P(k) = A k^{\gamma}$, where $k$ is the spatial frequency. The 3D power-law index $\gamma$ represents the relative strength of density fluctuations in different spatial frequencies.
When $\gamma$ is fixed, the amplitude $A$ is mathematically related to the width of the density PDF ($\sigma_{\rm s}$) based on Parseval's theorem. Therefore, in this study, we use the two parameters $\gamma$ and $\sigma_{\rm s}$ (instead of $A$) to parameterise clumpy density structures of clouds.

We use the SPS analysis toolkit {\sc Turbustat} Python package (\citealt{Koch_2019}) to identify  $\gamma$ from the HI image. We estimate $\sigma_{\rm s}$ by fitting a log-normal distribution to the column density PDF. The resulting $\gamma$ is -3.0 and $\sigma_{\rm s}$ is 0.24. 
Although these parameters are direct measurements of the Smith Cloud, we emphasize that the purpose of this paper is not to conduct a case study of the Smith Cloud. Discussions in this paper are rather applicable to any clumpy clouds moving at a high velocity with respect to the surrounding gas.

\begin{figure}
    \centering
    \includegraphics[width=\columnwidth]{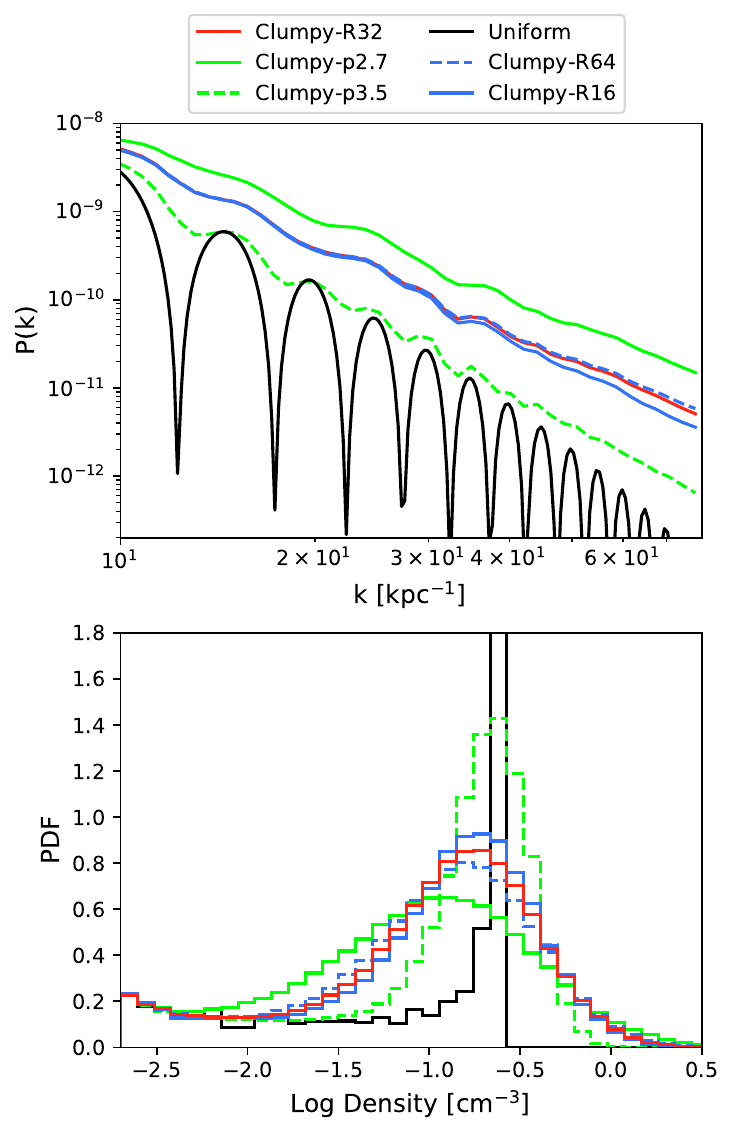}
    \caption{
    Top panel: The initial SPS of the clouds used in this study. Note that the amplitude $A$ is fixed when varying the spectral slope of clumpy clouds. The changes in resolution only marginally affect the spectra. The SPS of the Uniform cloud is simply a sinc function.
    Bottom panel: The initial density PDF of the clouds. 
    Compared to the fiducial Clumpy-R32 cloud (red line), the Clumpy-p2.7 cloud (green solid line) has a wider PDF that extends to a higher density and the Clumpy-p3.5 cloud (green dashed line) has a narrower PDF and a lower maximum density. The effect of resolution on the PDF is insignificant.
    }
    \label{fig:init_hist}
\end{figure}

We generate a 3D array of an overdensity ($\delta = \rho/\left<\rho\right> -1$) that follows the statistics ($\gamma$ and $\sigma_{\rm s}$) obtained from the above procedure. Although the reference $\gamma=-3$ is extracted from a 2D column density of the HI image, we directly take this value to build the 3D density structure. We confirm that the spectral slope of the clumpy cloud projected to 2D is within a reasonable range suggested by observations of the ISM though the exact value depends on the projection axis. In addition to $\gamma=-3$, we also build clouds with $\gamma=-2.7$ and $-3.5$ for the initial conditions of the MC-p2.7 model and the MC-3.5 model. The power spectrum amplitude $A$ is fixed in all cases, as a result, the change in $\gamma$ modifies the power to a greater extent on smaller physical scales.
See the top panel of Fig. \ref{fig:init_hist} for the initial SPS of all the clouds used in this study.

We use the {\sc powerbox} Python module (\citealt{Murray_2018}) which generates a periodic 3D scalar field with a given SPS. 
We first generate a cube at the highest resolution adopted in this work (i.e., 64 cells per cloud radius). We fix the spatial frequency $k=i/L$ for $i\in(-N/2, N/2)$, where $L=1.25\,\rm kpc$ and $N=800$, thus the output retains information about the assumed SPS at length scales larger than the highest resolution of the simulations. 
Then we degrade the 3D overdensity array to lower resolutions for models that use Clumpy-R32 or Clumpy-R16 as the initial condition. By doing so, we ensure that the density distributions at different spatial resolutions are fundamentally the same structures. 

Next, we apply a spherical kernel to the overdensity cube to set a boundary of the cloud that smoothly fades into the halo. The profile of the kernel is described as a function of the distance from the kernel centre (r):
\begin{equation}\label{eq:1}
f(r) = \frac{1}{2}(n_{\rm cloud}-n_{\rm halo})\left(1-\tanh \left[s\left(\frac{r}{r_{\rm cloud}}-1\right)\right]\right),
\end{equation}
where $n_{\rm cloud}$ is the core number density of the cloud,  and $n_{\rm halo}$ is the gas number density in the halo.
The steepness of the transition between the cloud and the surrounding medium is controlled by the parameter $s$. By definition, $r_{\rm cloud}$ ($=0.1\,\rm kpc$) is a radius where the kernel density is halfway between $n_{\rm cloud}$ and $n_{\rm halo}$. We set a relatively sharp kernel edge ($s = 30$) that corresponds to a transition between the cloud and the halo gas over $\sim 4$ cells at the standard resolution (i.e., 32 cells per cloud radius).
We move this spherical kernel around the overdensity cube until the centre of the mass of the structures within the kernel matches the kernel's centre within the size of one cell. This step is to prevent any imbalance of the mass distribution within a clumpy cloud.
We have confirmed that applying the spherical kernel has a negligible effect on the initial SPS at scales relevant to our study.
Once the kernel is multiplied to the overdensity field, $n_{\rm halo}$ is added to all the cells. Finally, the cube is plugged into the simulation as an initial density field. 
For all the variations of initial clumpy clouds used in this study, we scale the densities of the clouds so that their total cold gas masses are roughly the same. See Table \ref{tab:init} and discussions below for comparisons of the clouds.

For the models with a spherically symmetric initial cloud density profile (MU and HU in Table \ref{tab:t1}), we adopt a density profile that follows equation \ref{eq:1} with the steepness parameter $s = 10$. 
In this case, the cloud has a flat, uniform-density core and its edge smoothly fades into the surrounding halo gas.
Hereafter, we refer to this cloud structure as a ``uniform density'' cloud.

\begin{figure*}
    \centering
    \includegraphics[width=\textwidth]{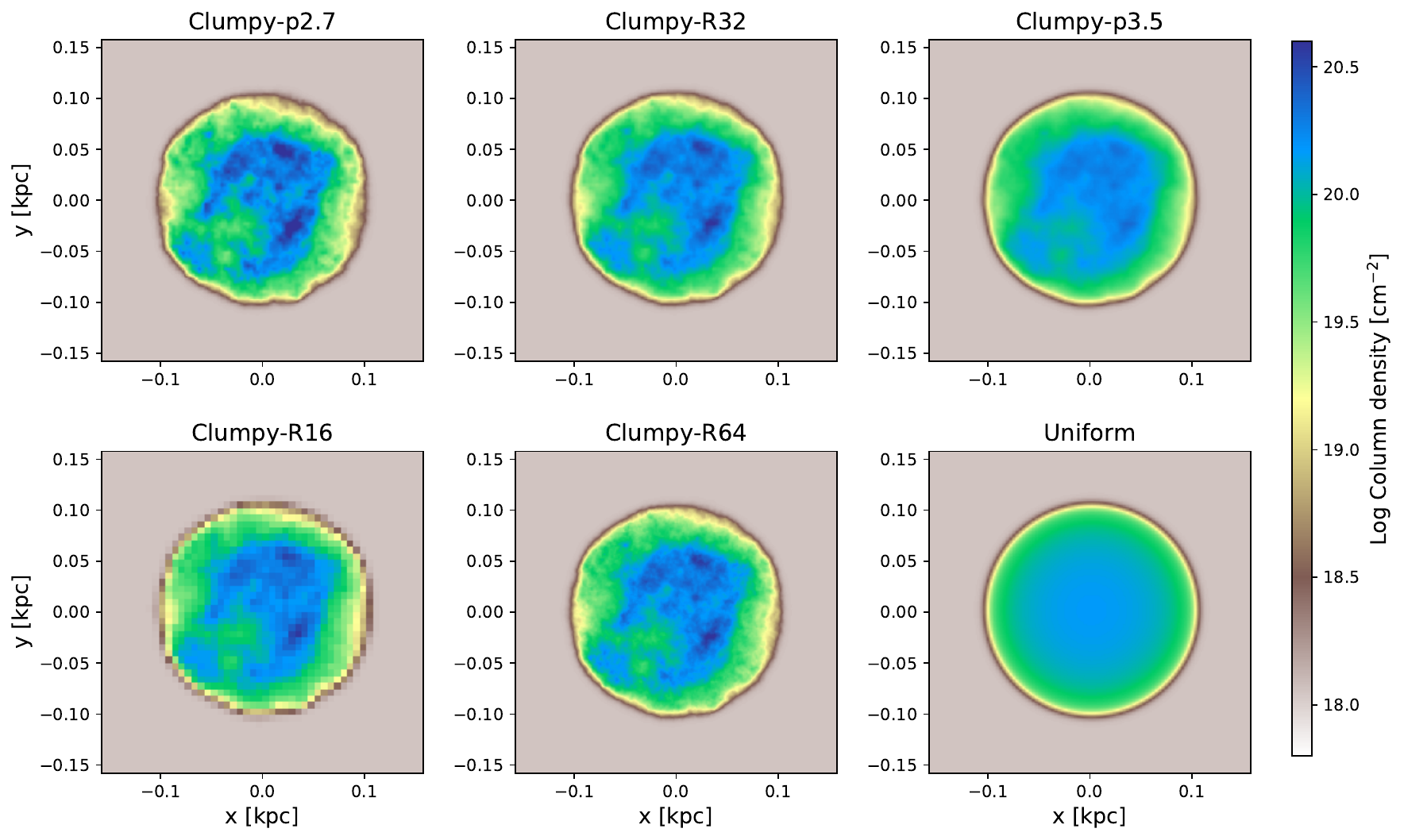}
    \caption{
    The column density of different initial clouds presented in Table \ref{tab:init}. The colour scale is fixed for all the panels for easy comparison between the panels. 
    }
    \label{fig:init_cloud}
\end{figure*}

In Fig. \ref{fig:init_cloud} we present the column density distribution of the clouds used in this work. The colour scale is fixed for all the panels for easy comparison between the panels.
The top three panels are the clouds with different power spectrum slopes $\gamma=-2.7, -3, $and $-3.5$ that we use for the MC-p2.7 model, the MC model, and the MC-p3.5 model, from left to right. 
As shown in the top panel of Fig. \ref{fig:init_hist}, we have kept the amplitude of the power spectrum fixed when varying the slopes.
As a result, a shallower power spectrum slope (Clumpy-p2.7) leads to a clumpier cloud with individual clumps highlighted and larger density contrast. 
In contrast, with a steeper slope (Clumpy-p3.5), the density contrast within the cloud is smaller and the clump structures are more diffuse.
In the bottom panels, we show the clouds for lower and higher resolution models, and the uniform density cloud.

The bottom panel of Fig. \ref{fig:init_hist} presents the density PDF of the initial clouds. The fiducial Clumpy-R32 cloud is shown in a red solid line. The black solid line marks the density PDF of the uniform-density cloud which peaks at the core density.
Depending on the power spectrum slope, Clumpy-p2.7 cloud (green solid line) has a wider distribution and Clumpy-p3.5 cloud (green dashed line) has a narrower distribution which is in line with the density contrast visible in Fig. \ref{fig:init_cloud}. 
The PDFs of the clouds with different resolutions are overall similar, with the distribution of the Clumpy-R64 cloud (blue dashed line) being slightly wider than that of the low-resolution Clumpy-R16 cloud (blue solid line).

As mentioned earlier, the cloud is initially in pressure equilibrium with the surrounding halo gas.
As a result, the log-normal density distribution of the cloud means that the temperature of the gas spans a certain range; a fraction of initial cloud gas cells are in temperatures above the cooling threshold ($T>10^4\,\rm K$, see Section \ref{sec:simul_overview} and Section \ref{sec:caveat}). 
With radiative cooling implemented, this gas could cool down to lower temperatures during the first few million years of the simulation.
We test the stability of the initial density field by letting the fiducial Clumpy-R32 cloud evolve with zero velocity relative to the surrounding halo gas. Because the initial mass budget in the intermediate-temperature cells is low ($\sim 3.5\%$ of the initial mass), the change in the density structure is negligible. The cloud stays stable throughout the time scale of the simulations  without any external influence.

In order to keep track of the advection of the material originally belonging to the cloud, we introduce a passive scalar tracer $C$ that advects with the flow: 
\begin{equation}\label{eq:tracer}
    \frac{\partial (\rho C)}{\partial t}+\nabla\cdot(\rho C \boldsymbol{v})=0.
\end{equation}
Initially, $C$ is set to 1 within the radius $r_{\rm cloud}=0.1\,\rm kpc$ from the cloud centre and 0 elsewhere.
In the following sections, we use $C$ as an indicator of the degree of mixing between the cloud and the halo gas within a certain cell.

Table \ref{tab:init} summarizes the properties of the initial clouds.
The second and the third column is the density ratio between the cloud and the halo gas, $\chi_{\rm mean}$ and $\chi_{\rm median}$, respectively, calculated by adopting the mean and the median density of all gas cells with $C=1$ as a representative density of a cloud.
In the fourth and the fifth column, we present the mass of the cloud calculated following two different definitions: (i) the mass of the cold phase gas ($T<2\times10^{4}\,\rm K$) and (ii) the mass within $r_{\rm cloud}=0.1\,\rm kpc$ from the clouds' centre where the passive scalar tracer $C$ is set to 1.
The difference between the two masses is minimal since much of the clouds' mass within $r_{\rm cloud}$ exists in the dense cold phase.

\subsection{Clump identification}\label{sec:clump_identify}

In this paper, we frequently refer to ``clumps'' in a cloud. 
These clumps are dense structures identified using the FellWalker algorithm (\citealt{Berry_2015}) implemented in the {\sc CUPID} package (\citealt{Berry_2007}).
The algorithm identifies clumps based on the gradients in the 3D density distribution.
We use any cells with densities higher than twice the density of the halo gas for the clump finding.
If a dip connecting two clumps is smaller than 30 times the halo gas density, the two clumps are considered as one.
We reject clumps with smaller than four cells along one of the axes in order to avoid clumps that are marginally resolved given the spatial resolution of the simulations.
We refer interested readers to \citet{Berry_2015} for detailed descriptions of the algorithm.
The total number of individual clumps identified is on the order of dozens in the fiducial MC model when it is reasonably evolved from the initial condition. 
The number changes over time and models. 

\section{The evolution of total cold gas mass}\label{sec:survivability}

\begin{figure*}
    \centering
    \includegraphics[width=0.9\textwidth]{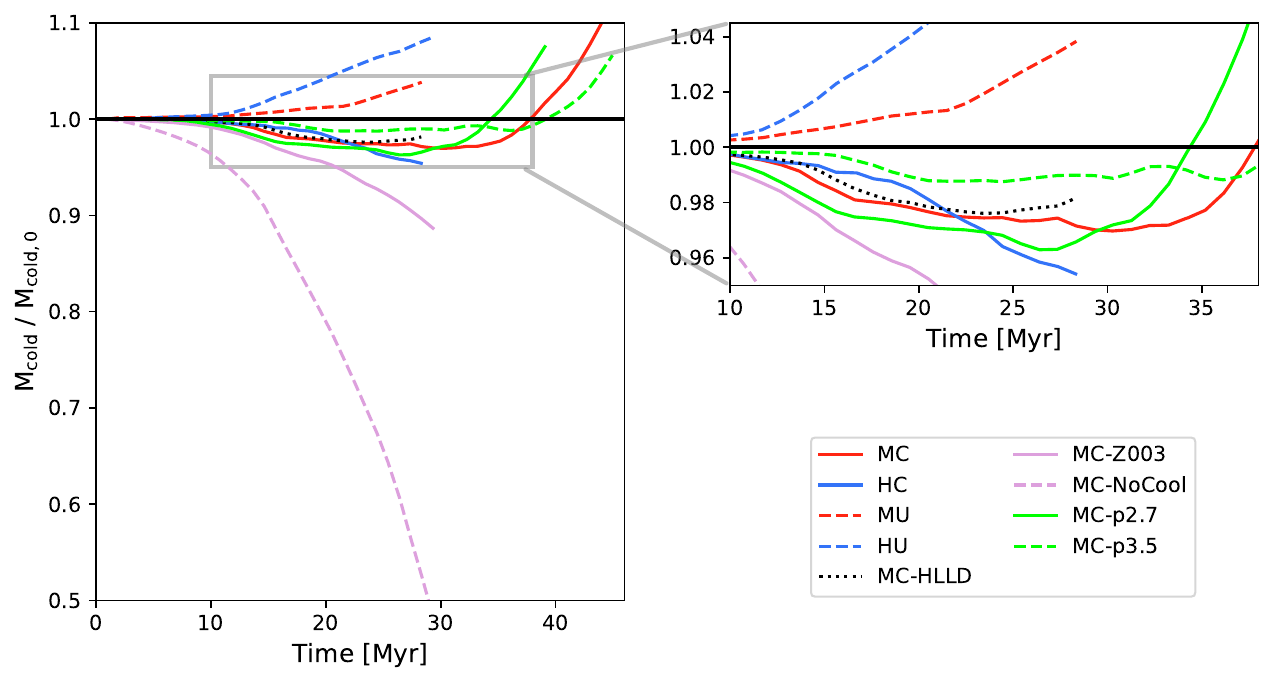}
    \caption{
    The evolution of the total cold gas mass in the simulations. The gas mass is normalized by the mass at $t=1\,\rm Myr$ (see the relevant paragraph of the main text for justification of this choice). The right panel is a zoom-in of $0.95<M_{\rm cold}/M_{\rm cold, 0}<1.05$ range for easy comparison between the models
    }  
    \label{fig:time}
\end{figure*}

The survivability of a cloud highly influences the question of whether or not a high-velocity cloud is capable of delivering cold gas to the Galactic disk.
If the radiative cooling of hot gas is faster than the cloud disruption, the total mass of the cold gas ($T<2\times10^{4}\,\rm K$) in the system grows over time. If not, the cloud eventually dissipates into the hot halo and no cold phase gas remains.
A characteristic timescale of high-speed relative motion impacting a cloud in a halo is often parameterized using the cloud crushing timescale ($t_{\rm cc}$); the time for a shock induced from the initial collision between the cloud and the surrounding medium to travel through the radius of the cloud
\begin{equation}\label{eq:tcc}
    t_{\rm cc} = \frac{r_{\rm cloud}}{v_{\rm halo}}\sqrt{\frac{\tilde{\rho}_{\rm cloud}}{\rho_{\rm halo}}}.
\end{equation}
Much effort has been made in parameterizing the survival of the cloud using scale-free parameters, e.g., $t_{\rm cool}/t_{\rm cc}$, where $t_{\rm cool}=\frac{3nk_{\rm B}T}{2n_{\rm H}^{2}\Lambda}$ (\citealt{Gronke_2018}; \citealt{Li_2020}; \citealt{Sparre_2020}).
However, in this work, we stick to rather qualitative interpretations.
This is because the actual degree of impact of the cloud-halo interaction changes significantly over time as the initial clumpy cloud breaks into multiple smaller pieces.
Therefore, while $t_{\rm cc}$ can still be a useful guide for some models, we would not expect it to be strictly applicable as a fundamental timescale of the evolution when it comes to clumpy clouds as in that of uniform density clouds.

It is important to clarify that we discuss the survival of the clouds by the total amount of cold gas in the system in this section. By doing so, we are open to the possibility of the original cloud body being completely destroyed, while the mass of cold gas within a simulation domain can still grow due to the formation of cold clumps condensed from the hot halo gas. 
This method is suitable for determining whether cloud systems moving at high velocity are capable of delivering gas to central galaxies, regardless of whether the gas is from the original cloud or not.
Similar considerations are often made in numerical studies that model cold ISM clouds under fast-moving hot galactic outflows (e.g., \citealt{Gronke_2018}; \citealt{Banda-Barragan_2021}) or cold ISM clouds launched into the halo as a part of the galactic fountain (e.g., \citealt{Armillotta_2016}).

Fig. \ref{fig:time} shows the evolution of the total mass of cold gas normalized by the mass at $t=1\,\rm Myr$ for easy comparison between the models. The change in the cold gas mass between the initial ($t=0\,\rm Myr$) and the first snapshot ($t=1\,\rm Myr$) is mostly due to the cooling of unstable gas in the initial condition. The amount of cooling is negligible, i.e. less than 2\% in all the models, but not zero. To exclude this effect, we normalize the cold gas mass with the gas mass at $t=1\,\rm Myr$. Due to the limited computing resources available, not all models are covered up to the time when the upturn of the cold gas mass might take place. For those models, discussions in this work are limited to the initial disruption phase of clouds.

We first focus on the overall evolution of the fiducial clumpy MC model (red solid line). 
Between 1 to $30\,\rm Myr$, there is a slight (3\%) decrease in the gas mass which is then followed by a dramatic growth from $35\,\rm Myr$. 
We confirm that the initial decrease in the mass is mainly due to the disruption of the cloud's main body, especially, the stripping of the lower-density interclump medium.
The later growth is due to the condensation of the intermediate-temperature gas that forms as a result of mixing in the wake behind the cloud. Further discussions on this topic will be presented in the following sections where we take a detailed look at the structural properties of the clouds.

Uniform density clouds (MU and HU; red and blue dashed lines) are always in the growth regime throughout the time covered by the simulations regardless of the presence of magnetic fields (MU: 104\% and HU: 108\% of the initial mass at $t\approx28\,\rm Myr$). 
In comparison, the mass of clumpy clouds (MC and HC; red and blue solid line) decreases during the same time frame.
We explain this discrepancy comes from the different density composition between the uniform/clumpy clouds (see the bottom panel of Fig \ref{fig:init_hist}):
the clumpy cloud has individual overdensities (i.e. clumps) of much higher densities compared to the uniform-density cloud, but its lower-density interclump gas is susceptible to ram pressure stripping and is therefore effectively removed at the cloud disruption phase.

Now we compare the evolution of the magnetized (MC and MU; red solid and dashed lines) and non-magnetized (HC and HU; red solid and dashed lines) models.
In the uniform cloud cases, the cold gas mass of the non-magnetized HU cloud is always higher than the magnetized MU cloud and the separation between the two curves grows with time. These results agree with previous studies with similar settings, e.g., \citet{Gronnow_2018} and \citet{Kooij_2021}. 
The role of magnetic fields in suppressing the growth of clouds is often discussed and interpreted as an outcome of inhibited Kelvin-Helmholtz mixing at the cloud-halo interface (\citealt{Gronnow_2018, Gronnow_2022}).
We show evidence of suppressed hydrodynamic instabilities in the presence of ambient magnetic fields in Section \ref{sec:mhd_vs_hd}. The difference between the HC model and the MC model is small throughout the time domain ($<1.8\%$). However, the decrease of mass in the MC model slows down and reaches a plateau by $t = 28\,\rm Myr$, whereas the HC model does not show a such trend. 
Yet, the time domain covered by the HC model is limited to the early phase of the evolution of the cloud and it is hard to predict the fate of the HC cloud at later times; the cloud may keep losing cold gas, but it could also turn around later like most of the other models.

With reduced or no radiative cooling,
both the MC-Z003 model (bright purple solid line) and the MC-NoCool model (bright purple dashed line) show a significant decrease in the cold gas (MC-Z003: 89\% and MC-NoCool: 52\% of the initial mass at $t=28\,\rm Myr$). 
In Section \ref{sec:metal}, we will show how the changed radiative cooling efficiency alters the density structure and the flow pattern around the clouds.
The clear discrepancy among the MC model, the MC-Z0.03 model, and the MC-NoCool model hints at the possibilities of diverse evolutionary tracks of observed HVCs depending on their metallicity (see Section \ref{sec:obs_metal} for further discussions).

Comparing the MC model to the models with shallower (MC-p2.7; green solid line) and steeper (MC-p3.5; green dashed line) initial density power spectra, we learn that varying the slope affects (i) how significant the initial mass loss is and (ii) how early the upturn of the mass we see in the evolution of the clumpy cloud takes place. 
The MC-p2.7 model loses 3.5\% of the initial cold gas mass before it enters the growth phase at $t=26\,\rm Myr$. This is a faster gas loss and earlier upturn of the mass compared to the evolution of the MC model which loses 3\% of the mass during the first $30\,\rm Myr$ and then starts to grow in mass. The MC-p3.5 model, in comparison, shows only 1.5\% initial mass loss and has an even later upturn that starts at around $t = 36\,\rm Myr$.
We explain that this trend is determined by how fast the clouds' main body breaks up and mixes with the surrounding halo gas which leads to condensation of halo gas in the wake. See Section \ref{sec:slope} for further explanations. 

\section{Comparison of clouds evolved in various environments}\label{sec:3}

\begin{figure*}
    \centering
    \includegraphics[width=0.9\textwidth]{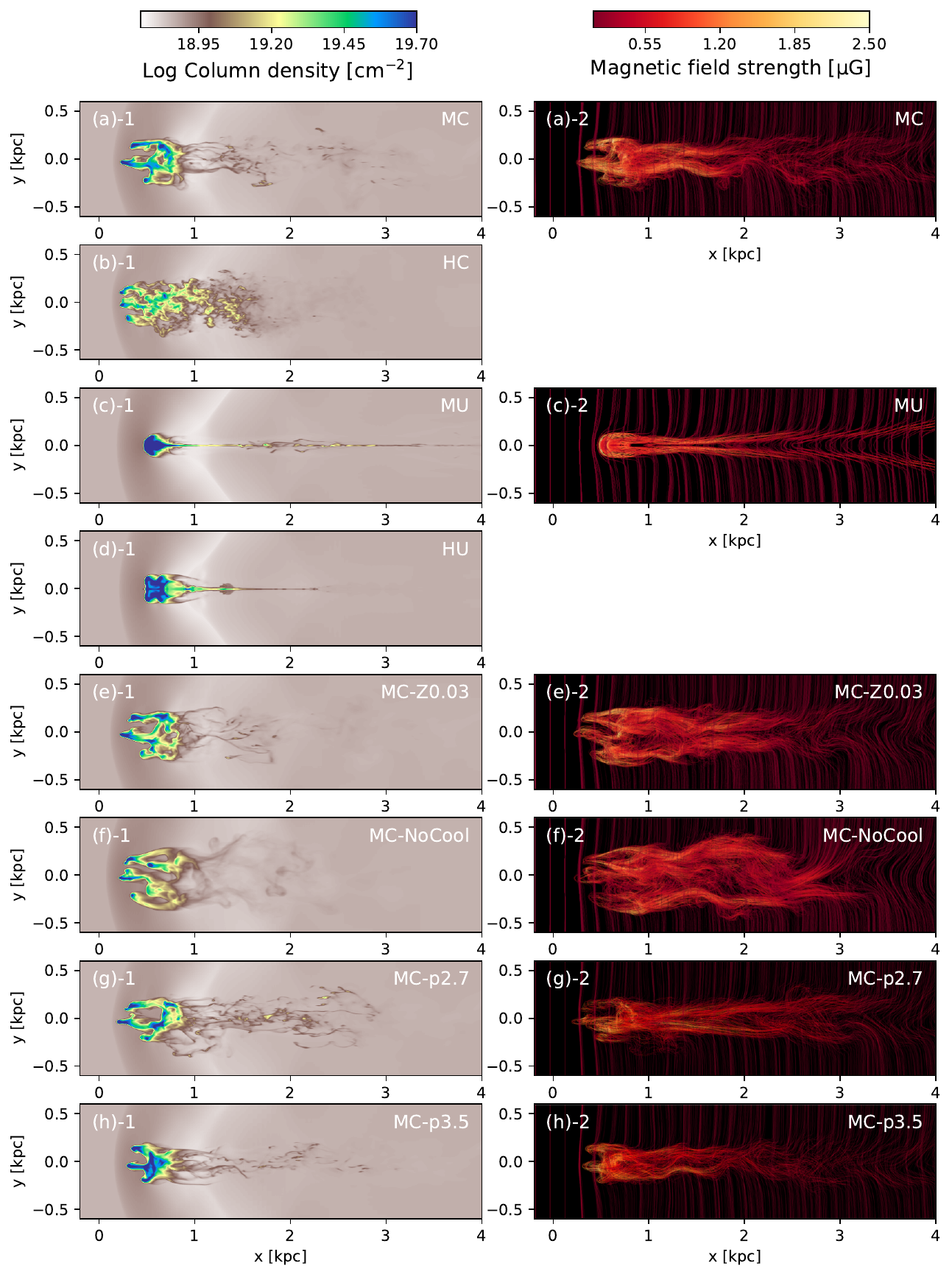}
    \caption{
    The density (left column) and the magnetic field (right column) distribution in different models. The model name is shown in the right corner of each panel. The left column shows the column density distribution integrated along the $z$-axis. In the right column, we show magnetic field lines coloured by the field strength along the lines. See the related texts for details of how the field lines for these figures are generated using VisIt software.
    }  
    \label{fig:all_model}
\end{figure*}

Interesting insights come from comparing details between the models.
By comparing the structural properties of clouds that started from identical/similar initial conditions but evolved in different environments, we can learn about what shapes the clouds' density structure throughout their evolution.
The evolution of the clouds is mainly governed by the following processes, which will hence be the focus of our interpretation of the results.
\begin{enumerate}
    \item \textit{Ram pressure} pushes gas towards the opposite direction of the cloud's motion. Given the high relative velocity between the halo gas and the cloud ($v_{\rm halo} = 200\,\rm km s^{-1}$), all our models are initially in the regime where ram pressure stripping is highly effective. 
    While ram pressure simply slows down the overall cloud motion in the case of uniform-density clouds, it can actually remove low-density gas from clumpy clouds in our models since gas in clumpy clouds spans a range of densities as shown in the bottom panel of Fig. \ref{fig:init_hist}.
    Further analysis and discussions are presented below in Section \ref{sec:fiducial_cloud}.
    
    \item \textit{Kelvin-Helmholtz and Rayleigh-Taylor instabilities} develop across the cloud-halo interface, leading to erosion of clouds from their edges.
    Such instabilities grow differently with and without the presence of magnetic fields, as magnetic tension provides extra stability to the mixing surface (\citealt{Chandrasekhar_1961}).
    This topic will be discussed in Section \ref{sec:mhd_vs_hd} where we perform a thorough comparison between the MC model and the HC model.
    
    \item \textit{Radiative cooling} takes place as cold cloud gas mixes with hot halo gas.
    The cooling rate ($\Lambda$) is a strong function of gas temperature and density that peaks at temperatures around $10^{5}\,\rm K$. This intermediate temperature gas is unstable and quickly cools down to lower temperatures $\sim 10^{4}\,\rm K$ and dense clumps can form in the wake behind the cloud's main body (see, e.g., \citealt{Marinacci_2010}; \citealt{Armillotta_2016}; \citealt{Gronke_2020}; \citealt{Banda-Barragan_2021}).
    Also, $\Lambda$ is roughly proportional to the gas metallicity ($Z$). In Section \ref{sec:metal}, we examine the effect of cooling on the cloud evolution by comparing the MC model ($Z = 0.3Z_{\rm \odot}$) with its lower metallicity counterpart MC-Z0.03 model ($Z \approx 0.03Z_{\rm \odot}$) and the MC-NoCool model with cooling turned-off.

    \item \textit{Shattering} is a physical process potentially important in understanding the evolution of fast-moving clouds (\citealt{McCourt_2018}) but not discussed in depth in this paper. See Section \ref{sec:resol} for further explanation.

\end{enumerate}
Although we have listed each of the above processes separately, it is worth mentioning that all of them act hand in hand. For example, as ram pressure pushes relatively cooler and lower-density cloud material, hydrodynamic instabilities induce mixing at the interface of the stripped material which leads to active radiative cooling in the region behind the main body of a cloud.

In this section, we focus on the snapshot at $t\approx 28\,\rm Myr$, which corresponds to approximately $4t_{\rm cc}$ since the onset of the simulations. See equation \ref{eq:tcc} for the definition of $t_{\rm cc}$.
That said, we do not impose any further meaning on $t_{\rm cc}$ than a rough estimate of the model environment; we do not attempt to, for example, utilize it as a quantitative measure of the evolutionary stages of the clouds.

Fig. \ref{fig:all_model} displays all models performed in this study at a fixed time of $t=28\,\rm Myr$\footnote{Links to movies showing the full time evolution of the clouds are provided at the end of the manuscript.
}. The left column (from panel a-1 to h-1) shows the column density distribution integrated along the $z$-axis (i.e., the axis perpendicular to the halo magnetic field and the direction of the cloud's motion).
The right column (from panel a-2 to h-2) is the 3D distribution of magnetic field lines coloured by the field strength at a given location. All panels are viewed along the $z$-axis.
These images are generated using VisIt software (\citealt{VisIt}).
For visualization purposes, we show only the field lines sampled from a set of seed locations of a $40\times3\times20$ mesh grid bounded within a box between $x = [-0.9,5.9]\,\rm kpc$, $y = [-0.5,0.5]\,\rm kpc$, and $z = [-0.8,0.8]\,\rm kpc$.

Models that started from the clumpy cloud initial conditions (MC, HC, MC-Z003, MC-NoCool, MC-p2.7, and MC-p3.5) share broad similarities in their cloud morphology: the main cloud body has three prominent ``heads'' and there are multiple filamentary tails extended along the direction of the motion.
In comparison, models with uniform density initial clouds (MU and HU) show a razor-thin tail.
Earlier studies have shown that such an extremely thin plane-shaped tail is an artefact of the spherically symmetric initial conditions for the cloud. When the perfect symmetry is broken by, for example, random internal velocities (\citealt{Kooij_2021}) or a random internal magnetic field (\citealt{McCourt_2015}), such a tail does not form even behind uniform density clouds.

\subsection{The fiducial model (MC)}\label{sec:fiducial_cloud}

\begin{figure*}
    \centering
    \includegraphics[width=\textwidth]{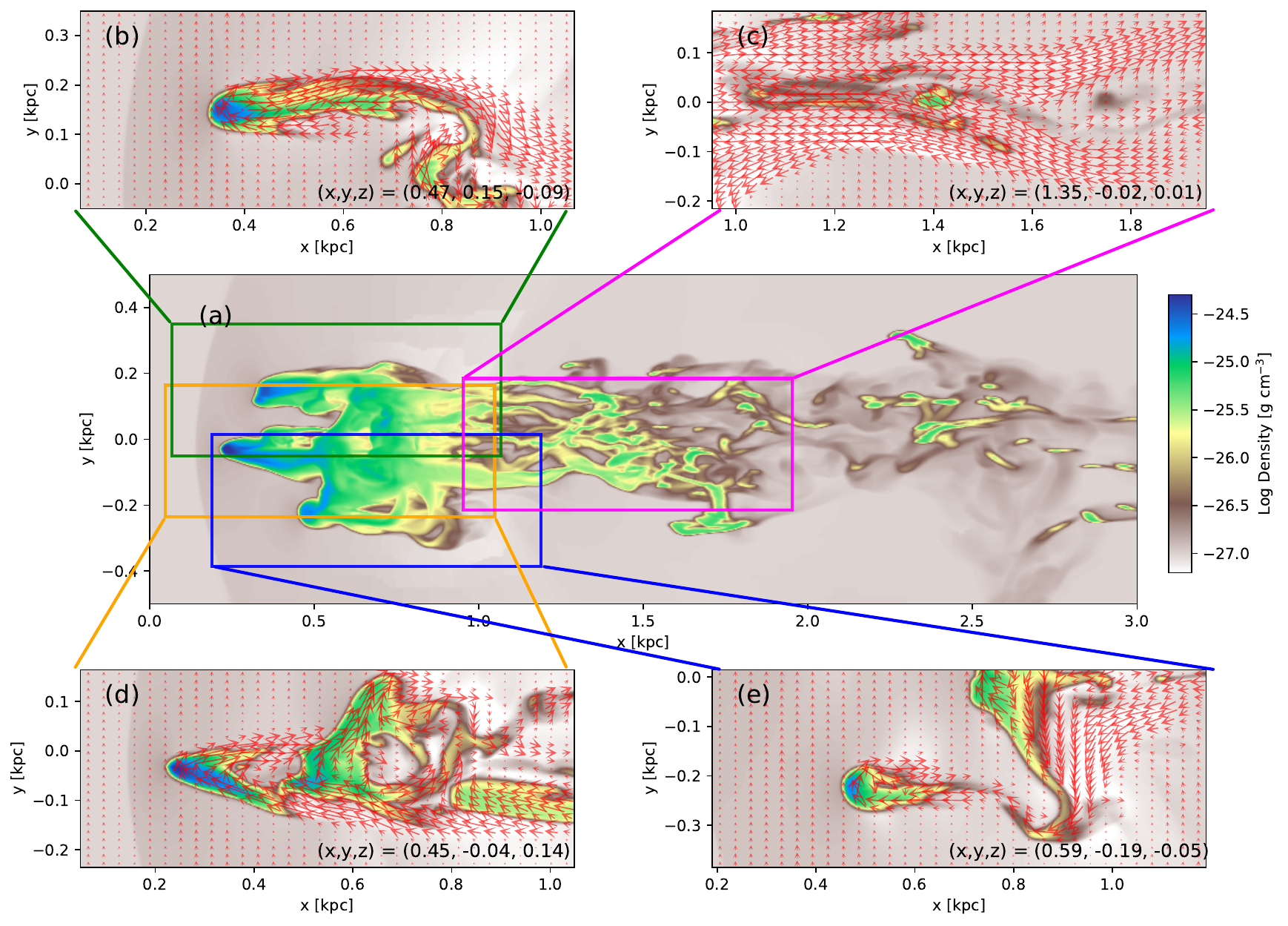}
    \caption{
    Panel (a): the density distribution of the MC model. The density presented here is the maximum density along the projection ($z$-axis). Panel (b), (c), (d), and (e): the density and magnetic field (red arrows) slice along different parts of the MC cloud. The coordinates in the bottom right are the centre of the mass locations of the clumps shown in the panels. Each slice is taken along the centre of mass $z$-coordinate of selected clumps. The three (panels b, d, and e) of the four panels are showing the main heads of the cloud. In these regions of the MC cloud, clumps have a clear head-tail morphology and magnetic fields drape around the clumps. Panel (c) is showing clumps in the wake of the cloud. Clumps in this region are threaded along long filaments and magnetic fields are overall parallel to the cloud's motion.
    }  
    \label{fig:fiducial_cloud}
\end{figure*}

We start by taking a close look at the cloud in the fiducial MC model. 
Fig. \ref{fig:fiducial_cloud} shows the density and the magnetic field in four selected regions of the MC cloud.
The main panel (a) in the middle presents the overall density distribution of the MC cloud. Three of the four zoom-in panels (panels b, d, and e) are slices of each head of the cloud along the $x-y$ plane and panel (c) takes a slice at the cloud's wake where we find groups of small cloudlets and filaments. 
The (x, y, z) coordinate in the bottom right corner of the zoom-in panels shows the centre of mass location of the clumps presented in each panel.
The red arrows show the direction of magnetic field vectors on the (x, y) slice and the size of the arrows is proportional to the field strength. 

The slices along the heads (panels b, d, and e) share common features: there is a clear head-tail density distribution and the magnetic field lines are draped around the overdensities.
To a certain extent, one can say that each head acts similarly to what earlier studies have found using simulations of uniform-density spherical clouds.
On the other hand, the density and magnetic field slice in panel (c) show somewhat different features. Several clumps are threaded into long filaments that extend along the direction of the cloud's motion. The magnetic field does not react to individual overdensities, but rather follows a large-scale magnetic field configuration that drapes around the entire cloud system. Therefore, the magnetic field configuration in this downstream region is overall parallel to the cloud's motion and the magnetic field draping around individual clumps is not efficient.

\begin{figure}
    \centering
    \includegraphics[width=\columnwidth]{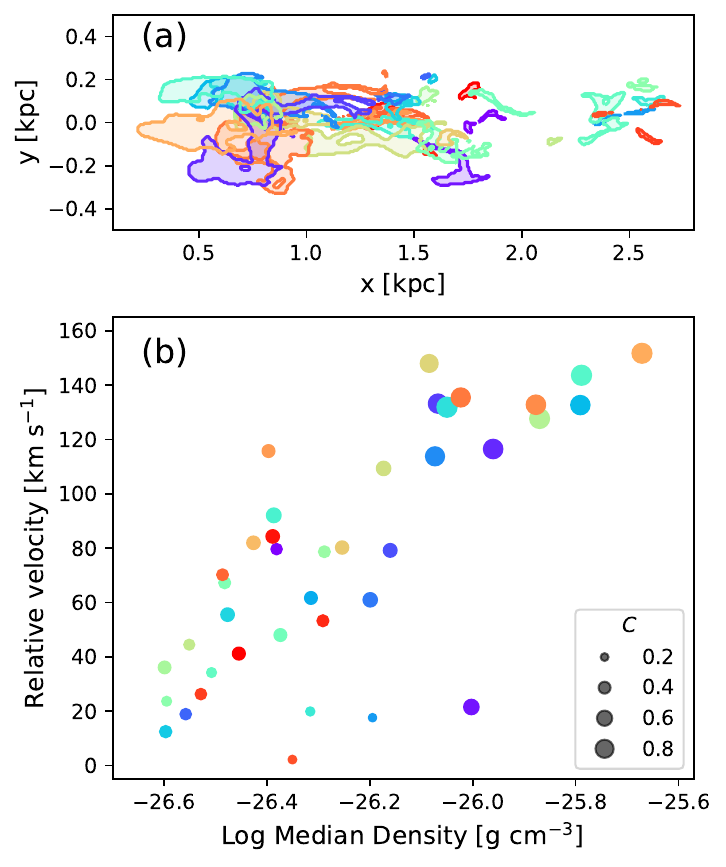}
    \caption{The properties of individual clumps in the MC model.
    Panel (a): The spatial distribution of clumps projected onto the $(x, y)$ plane. 
    Panel (b): the relative velocity of the clumps as a function of the median density. The size of the circles corresponds to the median passive scalar tracer ($C$) value within a clump, i.e., the fraction of the mass in each clump that was initially part of the cloud. Note the significant deceleration of clumps at larger distances from the main cloud body towards the downstream with lower densities and higher $C$ (i.e., a higher degree of mixing with the halo gas).
    }  
    \label{fig:fiducial_clumps}
\end{figure}

Using the clump finding algorithm described in Section \ref{sec:clump_identify}, we identify 39 clumps in this snapshot. See panel (a) of Fig. \ref{fig:fiducial_clumps} for their distribution on the (x, y) projected plane. In this panel, as well as panel (b), clumps are colour-coded differently for easy distinction.

In panel (b) of Fig. \ref{fig:fiducial_clumps}, we show the relative $x$-velocity between the clumps and the surrounding gas (i.e., $v_{\rm rel} = v_{\rm halo} - v_{\rm x, clump}$, where $v_{\rm x, clump}$ is the median $x$-direction velocity of a clump in the frame of the simulation domain) as a function of the median density of a clump.
We only consider the $x$-direction velocity here since the transverse velocity is almost negligible.
Each circle corresponds to a clump and the size of the circles is proportional to the median passive scalar tracer value within a clump ($0<=C<=1$) 
where $C=0$ means that all the cold gas has condensed from hot gas while $C=1$ means that all the cold gas was part of the initial cloud.
See equation \ref{eq:tracer} for the definition of $C$.
There is a clear deceleration throughout the extent of the cloud with the velocity gradient of about $130\,\rm km s^{-1}$ over $1.5\,\rm kpc$ across the $x$-direction.
The three head clumps have relative velocities around $140\,\rm km s^{-1}$. Compared to the initial velocity field where the halo gas is moving at $v_{\rm halo}=200\,\rm km s^{-1}$ and the cloud is static with respect to the simulation domain, these clumps are moving at $\sim70\%$ of their initial velocity and experiencing only half of the initial ram pressure along the $x$-axis considering that the magnitude of ram pressure is proportional to the square of the relative velocity ($P_{\rm ram} = \rho_{\rm halo}v_{\rm rel}^{\rm 2}$).

At the very low-velocity range (e.g., $v_{\rm rel}\lsim 20\,\rm km s^{-1}$), there are some clumps with higher median density than the overall density-velocity gradient. These clumps are located far behind the main cloud ($x\gsim2\,\rm kpc$) and their passive scalar tracer $C$ is around 0.2 or smaller, meaning that most of the mass comprising the clumps is from the surrounding halo gas. The ram pressure stripping and the growth of Kelvin-Helmholtz instability are almost negligible given their small velocity.
We explain the presence of these clumps is an outcome of the cooling of the halo gas onto material stripped from the cloud that acts as seeds for condensation.
The formation of a cooling-induced tail behind the main cloud is earlier discussed in \citet{Gronke_2018, Gronke_2020} where the authors find cold gas stripped from the main cloud converges and cools at the ``reformation point''.
In our models, the tail is not as spatially focused to one point because the perfect symmetry of the velocity field around the cloud is breached with the presence of the ambient magnetic fields and the clumpy initial density distribution. 
As for the low relative velocities of these clumps, we attribute the deceleration to the momentum transfer as a result of mixing between halo gas and the cloud fragments. Ram pressure and magnetic drag alone are not sufficient to decelerate the clumps as much within the time frame covered by the simulation. 
This explanation is in line with \citet{Banda-Barragan_2021} as well as the global galactic wind models of \citet{Schneider_2020} which demonstrate that the hydrodynamic mixing effectively alleviates the velocity difference between the hot outflowing gas and the cold ISM.
We continue detailed discussions on this topic in Section \ref{sec:metal} where we directly compare models with different radiative cooling efficiencies.

In reality, Milky Way's gravitational potential can further accelerate these low-velocity clouds/clumps towards the Galaxy.
Therefore, the external gravity field is required for simulations to examine the evolution and possible accretion of these condensed cold clumps in a more realistic sense.
One example of such work is \citet{Gronnow_2022} where the authors show that even with the gravity included many of the condensed clumps remain at low relative velocity with respect to the halo gas as the magnetic tension decelerates the clumps.
Such an effect has not been examined with clumpy clouds yet. We will postpone the investigation to future studies as it is beyond the scope of this work.

\subsection{Magnetic fields (MC vs HC)}\label{sec:mhd_vs_hd}

In this section, we focus on the effect of magnetic fields by comparing the fiducial MC model with its non-magnetized counterpart, the HC model. All other physical and numerical settings are identical in both simulations.
The density distribution of the MC cloud and the HC cloud is shown in panels (a)-1 and (a)-2 of Fig. \ref{fig:all_model}.
There are some similarities in the overall shape of the clouds that arise from having the same initial condition, e.g., the presence of the three main heads, however, clearly differences lie in their detailed structures.

\begin{figure*}
    \centering
    \includegraphics[width=0.8\textwidth]{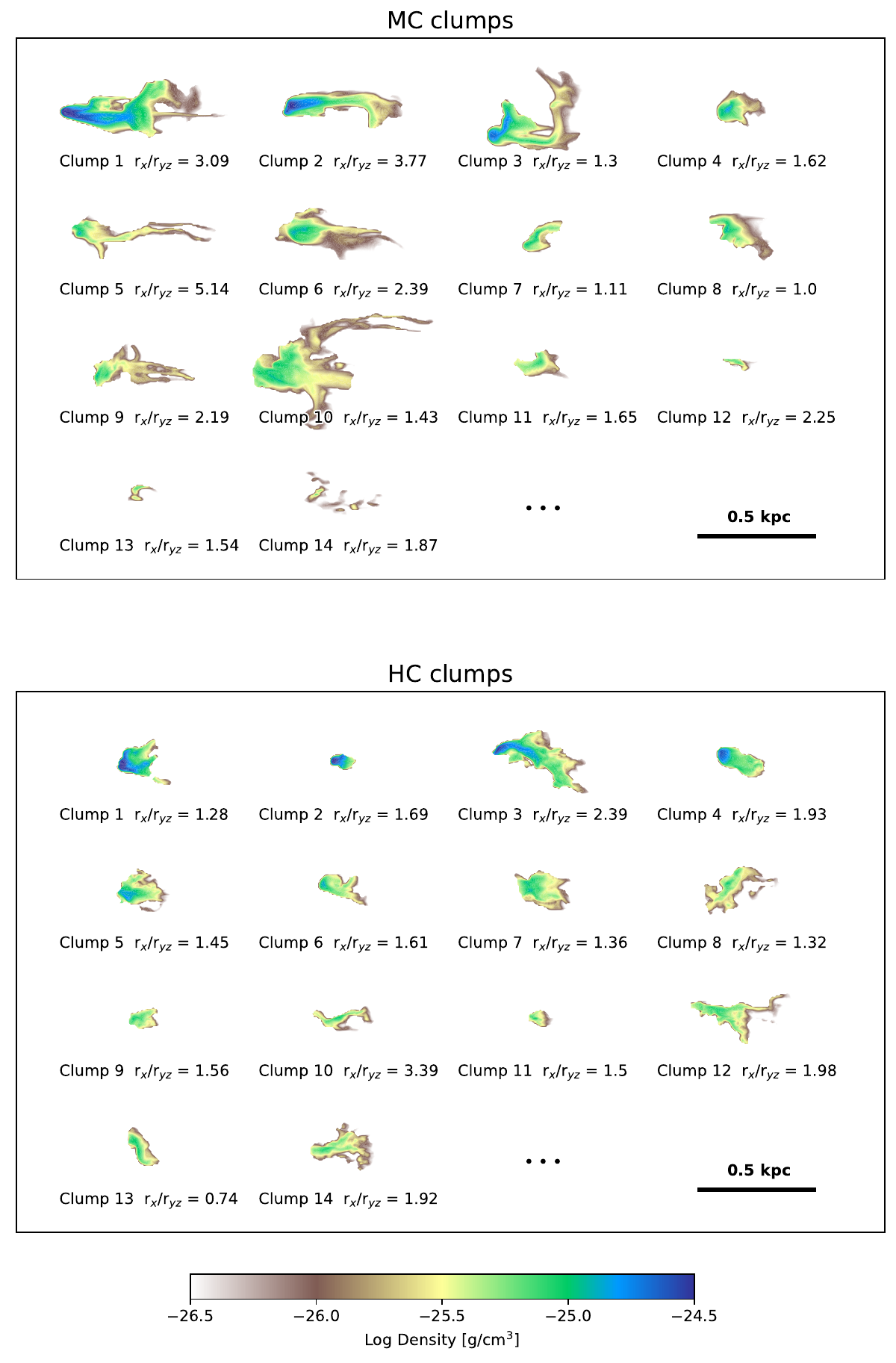}
    \caption{
    The density of the 14 densest clumps in the MC model (top) and the HC model (bottom).
    }  
    \label{fig:clump_comparison}
\end{figure*}

\begin{figure}
    \centering
    \includegraphics[width=\columnwidth]{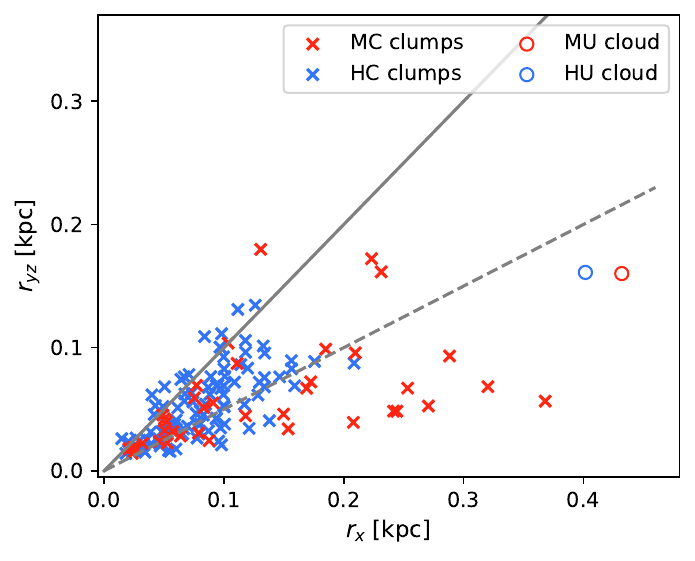}
    \caption{
    A comparison between the effective radius of clumps along the direction of cloud streaming motion ($r_{\rm x}$) and the axis perpendicular to it ($r_{\rm yz}=(r_{\rm y}+r_{\rm z})/2$). The red and blue crosses are clumps in the MC and the HC model, respectively. Only clumps above $100\,\rm M_{\odot}$ are presented.
    The effective sizes of the MU and the HU clouds are shown with circles. 
    We show $r_{\rm x} = r_{\rm yz}$ (solid line) and $r_{\rm x} = 2r_{\rm yz}$ (dashed line) relationships for reference. Overall, MC clumps are much more elongated compared to HC clumps due to suppressed hydrodynamic instabilities at the cloud-halo interface.}  
    \label{fig:clump_size}
\end{figure}

As shown in the previous section, there are 39 clumps in the MC model. With the same clump finding algorithm, we find significantly more clumps in the HC model: in total 101 clumps with in general smaller masses and sizes.
Furthermore, there is a notable difference in the overall shape of individual clumps in the two models.
In Fig. \ref{fig:clump_comparison} we show the 14 densest clumps identified in each model (top: MC, bottom: HC).
Clumps in the MC model are in general elongated along the $x$-axis, while HC clumps are fragmented into smaller sizes. The difference in the size of the clumps along the $y$-axis is not as significant, except for Clumps 3 and 10 in the MC model which have low-density tails notably extended along the $y$-direction.

In order to quantify the difference in clump sizes between the MC and the HC model, we measure the effective size of the clumps defined as 
\begin{equation}\label{eq:radius}
    r_{\rm i} = \sqrt{5\left(\left<X_{\rm i}^{2}\right>-\left<X_{\rm i}\right>^{2}\right)},
\end{equation}
where $i$ stands for each of the x, y, and $z$-axis and
\begin{equation}
    \left<X_{\rm i}\right>=\frac{\int X_{\rm i}\rho dV}{\int \rho dV}.
\end{equation}
Fig. \ref{fig:clump_size} shows the size of clumps along the $x$-axis ($r_{\rm x}$) to the clump size along the transverse axis\footnote{Earlier studies with spherically symmetric uniform-density clouds report that clouds often develop a flat tail in the $xy$-plane that is spread out along the $z$-axis ($r_{\rm y}<r_{\rm z}$) due to the enhanced growth of the RT instability along the axis perpendicular to the ambient magnetic field direction (i.e., $z$-axis in our models; see \citealt{Gregori_1999}; \citealt{McCourt_2015}; \citealt{Gronnow_2017, Gronnow_2018}). However, with the breach of the perfect symmetry in the MC model, such elongation is only prominent in a couple of large clumps while for most clumps taking the average of $r_{\rm y}$ and $r_{\rm z}$ holds as a reasonable representation of the size of clumps along the plane perpendicular to the motion.}, i.e., $r_{\rm yz}=(r_{\rm y}+r_{\rm z})/2$.
The red and blue crosses are clumps in the MC model and the HC model, respectively. We limit the sample points to clumps that are more massive than $100\,\rm M_{\odot}$.
For comparison, we present the effective size of the uniform density cloud models, MU (red circle) and HU (blue circle). The solid line and the dashed line show the $r_{\rm x} = r_{\rm yz}$ $r_{\rm x} = 2r_{\rm yz}$ relation, respectively, for reference.
The distribution of MC clumps is extended to larger $r_{\rm x}$ and larger $r_{\rm x}/r_{\rm yz}$ compared to that of HC clumps, again confirming what we have demonstrated from the qualitative comparison in Fig. \ref{fig:clump_comparison}.

The elongated shape of the MC clumps is not only simply an outcome of the streaming motion of the cloud but also a result of (i) a magnetic shield produced by magnetic field draping and (ii) the suppression of hydrodynamic instabilities at the cloud-halo interface.
As the cloud material advects downstream, clumps extend along the cloud's motion. 
Magnetic fields drape around each clump and produce a tube along the $x$-axis which pulls the cold gas along with it. 
This effect has been confirmed on the largest scale where a single uniform cloud will develop a long thin tail inside the draped tail field.
Also, when there are external magnetic fields in the halo, it provides extra tension at the cloud surface and stabilizes the perturbation. Therefore, clumps can further extend along the advection.
In contrast, albeit the HC cloud as a whole is overall extended along the cloud streaming direction (see panel b-1 of Fig. \ref{fig:all_model}) individual clumps in the HC model are shattered into smaller fragments rather than being elongated.
Any elongated structures can be easily perturbed without the presence of magnetic fields as hydrodynamic instabilities actively grow across the cloud-halo interface.

\begin{figure}
    \centering
    \includegraphics[width=\columnwidth]{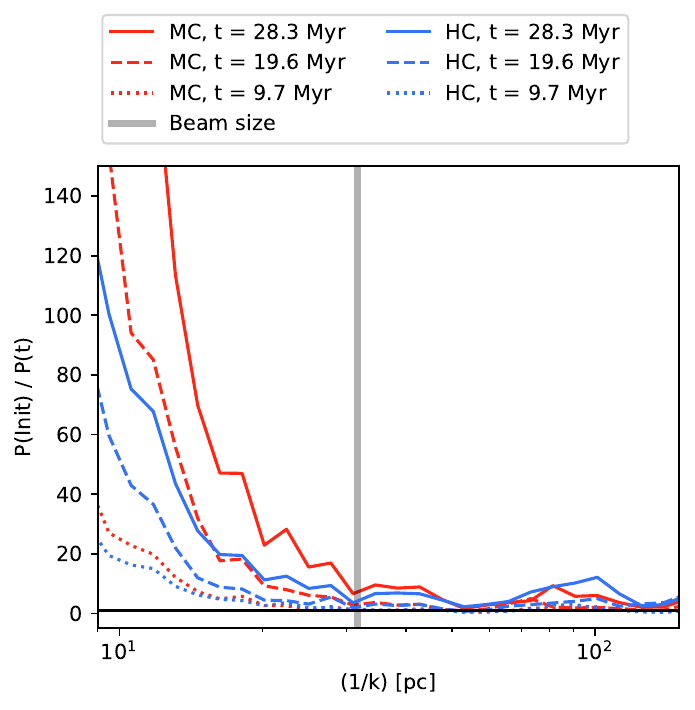}
    \caption{
    The evolution of the spatial power spectrum in the MC model (red) and the HC model (blue). The $x$-axis is the physical length scale ($1/k$) and the $y$-axis is the ratio between the initial SPS and the SPS at a given time $P(init)/P(t)$. The solid, dashed, and dotted lines show $t=28.3, 19.6,$ and $9.7\,\rm Myr$, respectively. The grey vertical line shows the observational beam size of the HI image from which the clump cloud's initial power spectrum was drawn. With time, $P(t)$ decreases significantly at small scales, while there is no systematic evolution at the scale larger than the beam size. At these scales, the HC model always has higher $P(t)$ than the MC model.
    }  
    \label{fig:time_power}
\end{figure}

For further quantitative analysis of the cloud structures in the MC model and the HC model, we compare the spatial power spectrum (SPS) of the models. 
We obtain the 2D SPS of the clouds by regridding all the AMR cells in the simulation domain to uniform-sized cells corresponding to the highest resolution and projecting the 3D volume along the $z$-axis to get the column density distribution.
We limit the analysis to cells with densities higher than twice the halo gas density ($>2\rho_{\rm halo}$). 
The 2D power-spectrum frequency scale is defined as $k = \sqrt{k_{\rm x}^{2}+k_{\rm y}^{2}}$, where $k_{\rm x}$ and $k_{\rm y}$ are the spatial frequencies along the $x$ and $y$-axes, respectively. In Fig. \ref{fig:time_power}, we show the evolution of the SPS for the MC model and the HC model.
The $x$-axis is converted to the physical length scale $(1/k)$ in units of pc. The smallest scale shown in the figure is $\sim 10\,\rm pc$, which is about three times the size of the maximum resolution cell of the simulations.
The vertical axis of the figure is $P(init)/P(t)$, where $P(init)$ is the median Fourier SPS amplitude at a given $k$ of the initial density distribution and $P(t)$ is this amplitude at a given time $(t)$. The black horizontal line is at $P(init)=P(t)$.
The red and blue solid lines show the SPS of the MC model and the HC model, respectively, at $t=28\,\rm Myr$.
Initially, at $t=0\,\rm Myr$, both the MC model and the HC model start from the same density power spectrum as described in Section \ref{sec:clumpy_setting}.
At $t=28\,\rm Myr$, both the models have less power at smaller length scales below $\approx 78\,\rm pc$ ($25\,\rm cells$) compared to the initial spectrum, i.e., higher $P(init)/P(t)$.
Furthermore, $P(init)/P(t)$ of the MC model is always higher than that of the HC model at this length range indicating that the growth of small-scale structures in the MC model is considerably suppressed in comparison with the HC model. 
We confirm that these trends are visible regardless of the projection axis, i.e., whether the 2D column density is integrated along the initial magnetic field direction ($y$-axis) or the axis perpendicular to it ($z$-axis).
For reference, we show $P(init)/P(t)$ at earlier times in dotted ($t=9.7\,\rm Myr$) and dashed ($t=19.6\,\rm Myr$) lines. 
In both the MC and the HC models, the difference between $P(init)$ and $P(t)$ grows over time to the direction where the power at the small scales significantly decreases.

Note that we construct the initial SPS of the clumpy cloud models based on observations of the Smith Cloud which is known to have interacted with the Milky Way over a period of $\sim 70\,\rm Myr$ (\citealt{Lockman_2008}; \citealt{Nichols_2009}). 
The gray vertical line in Fig. \ref{fig:time_power} shows the telescope beam size of the HI observation by \citet{Lockman_2008} that we draw the reference SPS from. The angular beam size is converted to the physical size using the distance to the Smith Cloud ($\sim 12.5\,\rm kpc$).
It is important to point out that there is no notable evolution of $P(t)$ at length scales larger than the beam. This result demonstrates that we can use the Smith Cloud’s SPS as a reference for the initial conditions of the simulation, even if we do not know about the structure of the HVC at the time the cloud started interacting with the Galaxy. In other words, constructing the initial condition based on the cloud already interacting with the Galaxy does not contradict the structural evolution of clouds demonstrated by our models.

\subsection{Radiative cooling (MC vs MC-Z0.03 vs MC-NoCool)}\label{sec:metal}

We study the effect of radiative cooling on the cloud's structural properties by comparing three models with different cooling efficiency:
the fiducial MC model with $Z=0.3Z_{\odot}$, the MC-Z0.03 model with $Z=0.03Z_{\odot}$ that represents the nearly pristine gas, and the MC-NoCool model without radiative cooling implementation. All other parameters are set the same. The lower metallicity of the MC-Z0.03 model compared to the MC model means less effective radiative cooling since the cooling rate is roughly proportional to metallicity.
In Fig. \ref{fig:all_model}, panel (a)-1, 2, (e)-1, 2, and (f)-1, 2 show the density and magnetic field distributions of the models at $t = 28\,\rm Myr$.

\begin{figure*}
    \centering
    \includegraphics[width=\textwidth]{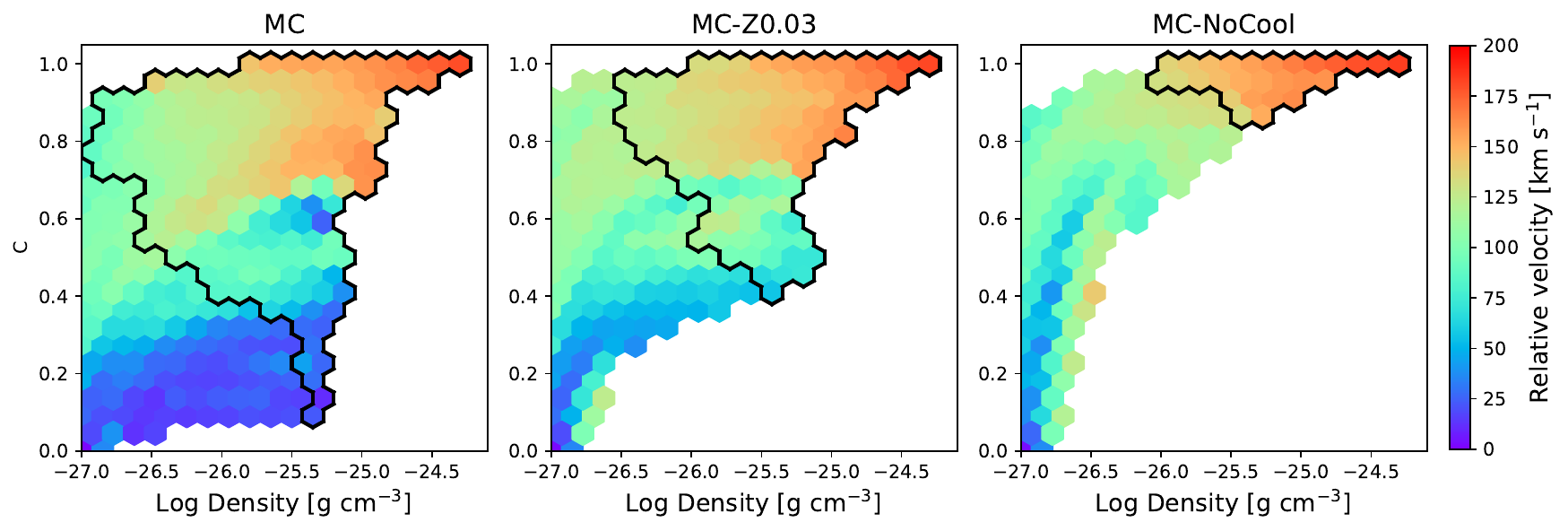}
    \caption{
    The distribution of gas on the density - passive scalar tracer $C$ plane, where $C$ is the mass fraction of the gas that was part of the original cloud. 
    The colour of each hexagon bin represents the mean $x$-direction relative velocity as defined in Section \ref{sec:fiducial_cloud}.  The black line encloses hexagon bins with cold gas ($T<2\times10^{4}\,\rm K$). Although not presented in the figure, note that the majority of the mass is distributed in the upper-right bins, i.e., dense original cloud material not yet mixed with the halo gas.
    In the MC model, there is dense gas that is significantly decelerated (purple colour) and mostly condensed from the halo gas (bottom right corner of the panel).
    In comparison, gas in the MC-NoCool shows a smooth transition from the dense clouds (top right) to low-density halo gas (bottom left). The MC-Z0.03 model shows characteristics between the MC and the MC-NoCool models.
    }
    \label{fig:cooling_tr}
\end{figure*}

In Fig. \ref{fig:cooling_tr}, we present the distribution of all gas cells in the MC model (left), the MC-Z0.03 model (middle), and the MC-NoCool model (right) on the density - scalar tracer $C$ plane, where $C$ is the fraction of the gas that was part of the original cloud and, therefore, a measure of the cold gas condensation (see equation \ref{eq:tracer} and the related paragraphs in Section \ref{sec:method}). The colour of each hexagon bin is the mean relative velocity ($v_{\rm rel}=v_{\rm halo}-v_{\rm x, clump}$) of gas cells within the bin.
The black line encloses the hexagon bins with cold gas ($T<2\times10^{4}\,\rm K$).
These plots need to be interpreted with caution as the mass in different hexagonal bins differs and the large majority of the mass is, in all three cases, contained in the dense remnants of the original cloud, i.e., in the upper-right bins.
The MC model has decelerated (purple colour) high-density gas cells that are mostly condensed from the halo (low $C$) as pointed out in Section \ref{sec:fiducial_cloud} and as in line with earlier models of, for example, \citet{Gronke_2018, Gronke_2020}.
In contrast, we do not see such a population of gas cells in the MC-NoCool model. Gas cells in the MC-NoCool model show a smooth transition from the high-density high-$C$ cloud material to the low-density low-$C$ halo gas.
This is because the material stripped out of the original cloud smoothly mixes and diffuses into the halo without any further cooling.
In the MC-Z0.03 model, radiative cooling exists but is not as efficient as in the MC model. Therefore, the gas cells' properties lie between the MC model and the MC-NoCool model.
This result aligns with the finding that we do not see as many dense clumps and filaments in the wake behind the clouds in the MC-Z0.03 model and the MC-NoCool model compared to the MC model. 
Such significance of the radiative cooling highlighted in our models is in good agreement with what is demonstrated in models of cold clouds launched into the halo via the galactic fountain process (e.g., \citealt{Armillotta_2016}) and models of cold ISM under fast-moving hot galactic outflows (e.g., \citealt{Gronke_2018}; \citealt{Schneider_2020}; \citealt{Banda-Barragan_2021}).

Furthermore, radiative cooling efficiency changes the compactness of the clouds. Less radiative clouds are more adiabatic and the heat they gain increases thermal pressure and causes expansion.
The distribution of magnetic field lines also follows a similar trend.
In panels a-2, e-2, and f-2 of Fig. \ref{fig:all_model}, we find that the field lines are much tighter, denser, and better aligned in the tail of the MC model than in the MC-Z0.03 model and the MC-NoCool model. Consequently, the field strength is also overall higher in the MC model as it is given by the density of field lines. The tighter field is due to the field lines being draped over the smaller surface of the more compact MC cloud.

\begin{figure*}
    \centering
    \includegraphics[width=0.95\textwidth]{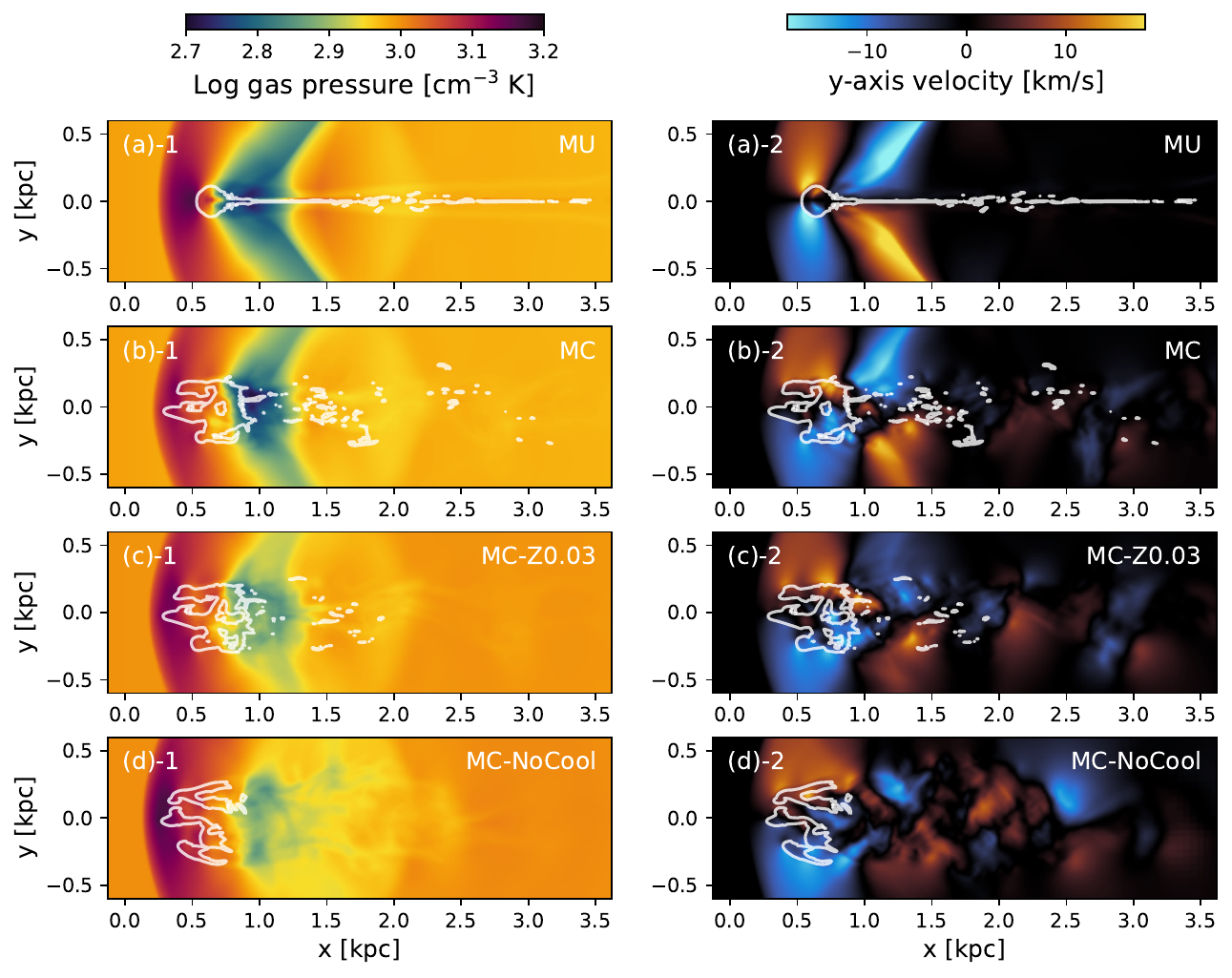}
    \caption{The average projected pressure (left) and $y$-velocity (right) around the clouds in different models. The $y$-direction is the axis parallel to the halo magnetic fields and perpendicular to the cloud's motion. The white contour shows the cloud's density distribution. 
    From top to bottom panels (with an exception of the MU model), radiative cooling becomes less efficient. As a result, the main body of the cloud becomes less compact, the pressure gradient in the front and behind the cloud is less steep, and the velocity field is more highly disturbed.
    }  
    \label{fig:flow}
\end{figure*}

Consequently, the change in the radiative cooling efficiency alters the overall flow pattern in and around the clouds.
Fig. \ref{fig:flow} presents the gas pressure (left panels) and the $y$-velocity (right panels) averaged along the $z$-axis. The white contours show the density distribution as shown earlier in Fig. \ref{fig:all_model}. Note that our choice to show the average properties along the projection axis is to present the \textit{overall} patterns of the pressure and the $y$-velocity. When the properties across a slice (e.g., $z=0$ slice) are considered, it is not as straightforward to compare the patterns between the models because the clouds in different models have different structures.

The pressure distribution in the MU model (panel a-1) shows a clear pressure gradient in the front and behind the cloud as a result of the shock compression due to the cloud's high-velocity motion. 
The $y$-velocity field (panel a-2) around the MU cloud has a flow pattern close to the quadrupolar distribution: the flow of halo gas at $y>0$ and $y<0$ separate at the cloud's head and converge behind the cloud. 
The pressure and the $y$-velocity patterns in the MC model (panel b-1 and b-2) are largely alike to that of the MU model, but with a slight distortion due to the breached symmetry. Such similarity indicates that the main cloud body of the MC model acts in a similar way to the one in the MU model in determining the flow around the cloud. 
On the contrary, we find noticeable differences in the flow pattern in models with no/less cooling (panel c-1, 2, and panel d-1, 2). 
The diverging flow along the $y$-axis at the cloud head is still noticeable, but the converging flow behind the cloud is not as coherent as in the MU and the MC models with higher metallicity. The presence of the low-pressure region behind the cloud is also not as obvious.

At the initial snapshot, dense cores in the clumpy cloud are embedded in the inter-clump gas that is denser than the halo gas and not as dense as the cores (see Fig. \ref{fig:init_cloud} and Fig. \ref{fig:init_hist}).
These intermediate-density materials are more susceptible to the ram pressure stripping and therefore the halo gas can easily penetrate the cloud along the narrow gap between the clumps.
When there is no/less radiative cooling in the system, such internal flow can easily reduce the pressure gradient and easily produce a disturbed flow pattern that continues to the wake behind the clouds.
In contrast, when radiative cooling is efficient, the intermediate-density inter-clump gas can easily cool down to a higher-density phase and act like bonds between the existing cores.
In this case, the halo gas faces stronger resistance in flowing through the cloud, so much of the flow goes around the cloud's outer surface instead.

In summary, the role of radiative cooling in cloud evolution is to (i) produce dense clumps and filaments in the wake behind the cloud and (ii) keep the main body more compact.

\subsection{Density power spectrum (MC vs MC-p2.7 vs MC-p3.5)}\label{sec:slope}

The slope of the density spatial power spectrum ($\gamma$) describes the relative strength of density fluctuations in different spatial scales. 
In this section, we discuss the effect of varying $\gamma$ on the evolution of clouds by comparing the fiducial MC model ($\gamma = 3$) to the MC-p2.7 ($\gamma = 2.7$) model and the MC-p3.5 ($\gamma = 3.5$) model. See the left panel of Fig. \ref{fig:init_hist} for the exact form of the initial power spectra of the models. While we vary the spectral slope, the amplitude of the spectra remains fixed.

The density and magnetic field distributions of the models at $t = 28\,\rm Myr$ are in panel (a)-1, 2, (g)-1, 2, (h)-1, 2 of Fig. \ref{fig:all_model}.
Regardless of the initial power spectrum slope, all the models develop the three leading head-like structures. 
However, the extent to which the heads are exposed from the main cloud body varies: in the MC-p3.5 model all the heads are still tightly attached to the trailing main cloud while in the MC-p2.7 model, they are well separated from each other.
Another clear difference we find with the change of the power spectrum slope is the amount of dense filamentary structures we see in the wake behind the cloud. The MC-p2.7 model has more prominent filaments and dense clumps compared to the MC and the MC-p3.5 models, which indicates efficient mixing of the cloud material and the halo gas and, as a result, more condensation via radiative cooling. Note that all parameters other than the power spectrum slope are fixed, including the metallicity and the radiative cooling efficiency.

The differences between the models come from the different density contrast \textit{within} the initial clumpy clouds, i.e., the width of the density PDF in the bottom panel of Fig. \ref{fig:init_hist}.
Earlier in Section \ref{sec:clumpy_setting}, we compare the structures of Clumpy-R32, Clumpy-p2.7, and Clumpy-p3.5 clouds that are put to use as initial conditions of the MC model, the MC-p2.7 model, and the MC-p3.5 model, respectively.
In short, the shallower the slope of the power spectrum is, the higher the density contrast within a cloud, i.e., the cloud has denser cores and less dense inter-clump material.
With time, such low-density inter-clump gas can easily be pushed away from the cloud body. Therefore, we see dense heads well separated from each other in the MC-p2.7 model with lower-density inter-clump material compared to the other models.
Furthermore, the displaced inter-clump gas 
mixes with the hot halo gas and condenses in the wake behind the cloud which explains why we find more dense filaments and clumps in the tail region of the MC-p2.7 model compared to the other models.
This effect is also noticeable in the cold gas mass evolution of the MC-p2.7 model, where we find a faster initial disruption followed by a significant growth of the mass (see Section \ref{sec:survivability} for detailed discussion).

\section{Discussion}

\subsection{Resolution effect}\label{sec:resol}

\begin{figure}
    \centering
    \includegraphics[width=\columnwidth]{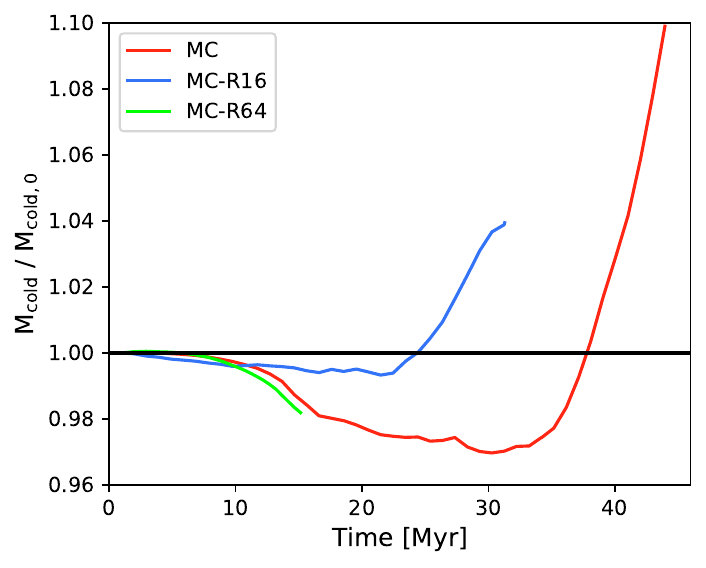}
    \caption{The same format as Fig. \ref{fig:time}, showing the evolution of models with different spatial resolutions.}  
    \label{fig:time_resol}
\end{figure}

To assess the effect of the simulation resolution, we ran simulations identical to the MC model, but with one less and one additional AMR level. This halves (MC-R16) and doubles (MC-R64) the maximum resolution, respectively. As can be seen in Fig. \ref{fig:time_resol}, the lower resolution simulation loses less cold gas and enters the growth regime earlier than at higher resolutions. It is overall more similar to the evolution of the uniform density cloud (MU model) than at a higher resolution. This is expected because the lower resolution smoothes out some of the density structures making the distribution closer to the uniform density cloud. We could only run the higher resolution case until $t\sim15\,\rm Myr$ due to its high computational cost, however, over the time domain covered, it quite closely follows our standard resolution MC model. It loses cold gas slightly faster than the MC model as expected given that the MC model also loses cold gas faster than the lowest resolution case. Given this, we expect MC-R64 to also eventually enter the growth regime but a bit later than the MC model. This is in line with the findings for the resolution dependence of cold gas growth for magnetized uniform clouds (\citealt{Gronnow_2018}).

\citet{McCourt_2018} demonstrate that any large cloud complex is susceptible to shatter into fragments as small as a characteristic length scale of $l_{\rm cloudlet}\sim c_{\rm s}t_{\rm cool}\sim 0.1 \,{\rm pc} /n \,{\rm cm^{-3}}$. In our models, the expected shattering length scale is about $0.5-1\,\rm pc$ which is smaller than the best resolution of the fiducial model (i.e., $3.125\,\rm pc$). Therefore, we postpone the investigation of the shattering process with the presence of magnetic fields and the initially clumpy density distribution to future studies.

\subsection{Caveats}\label{sec:caveat}

Observational evidence suggests that the ISM as well as the surrounding hot halo gas are turbulent (\citealt{Jansson_2012}; \citealt{Beck_2016}).
In this work, we simulate a cloud with a realistic clumpy-density structure motivated by observed HI clouds.
But in fact, not only is the clouds' density distribution clumpy but their velocity and magnetic field structures are also turbulent. 
A number of studies to date have taken into account the turbulent nature of the cloud-halo interaction to some degree, e.g., internal turbulent velocity fields of clouds (\citealt{Heitsch_2021}), tangled magnetic fields within clouds (\citealt{McCourt_2015}; \citealt{Gronke_2020}) and turbulent magnetic fields in the halo gas (\citealt{Sparre_2020}), or all of the above (\citealt{Banda-Barragan_2018,Banda-Barragan_2019}).
There is a clear indication that turbulent halo magnetic fields act somewhat differently to uniform fields when draping a cloud: they suppress Kelvin-Helmholtz instabilities in both directions that are perpendicular to the cloud's motion (\citealt{Sparre_2020}).
That said, at the very vicinity of the cloud-halo interaction front, random magnetic fields are stretched and flattened out and therefore act like a transverse, ordered field (\citealt{Asai_2007}) similar to the uniform halo magnetic field adopted in the models in this study.

We do not include such non-uniform velocities or magnetic fields in models in this study in order to isolate the effect of the clumpy density structure from that of velocities or magnetic fields as well as to keep the number of free parameters in the simulations manageable. In addition, the turbulent velocity and tangled magnetic fields of HVCs are not as well identified by observations as their density structure since these are more complicated and hard to observe.

Also, it is worth noting that many of the studies mentioned above typically aim to simulate a cloud being launched from a galactic disc or a static ISM cloud under fast-moving hot galactic outflows. In both cases, clouds originate from the turbulent galactic ISM environment. 
In comparison, the parameter space of our models represents HVCs at larger distances infalling from the intergalactic space. Thus far, it is not well studied to what extent such clouds are affected by the turbulent environment.

In this work, we do not simulate the metallicity difference between the cloud and the halo gas. We instead set a uniform metallicity throughout the simulation domain as previously mentioned in Section \ref{sec:method}.
\citet{Miller_2013} estimated a lower limit of the hot-phase halo gas metallicity to be $\sim 0.2\,\rm Z_{\odot}$ by comparing the electron density estimated from their OVII K$\alpha$ absorption line measurements and the dispersion measure of pulsars in the Large Magellanic Cloud (\citealt{Anderson_2010}).
Having said that, using the same metallicity for the halo and HVC is well motivated by the fact that the observed metallicities of the halo gas ($\sim 0.2\,\rm Z_{\odot}$) and HVCs ($0.1-0.3\,\rm Z_{\odot}$) largely overlap. 

Another physical process that may affect the evolution of cold clouds but is omitted in this work is thermal conduction.
Hydrodynamic simulations in general show that thermal conduction suppresses the growth of hydrodynamical instabilities at the cloud-halo interface (\citealt{Vieser_2007a, Vieser_2007b}; \citealt{Armillotta_2016, Armillotta_2017}; \citealt{Bruggen_2016}).
With the presence of magnetic fields, electrons, which are the main mediators of thermal conduction, travel along the magnetic field lines that are not necessarily in the direction that reduces the temperature gradient efficiently. Thus, thermal evaporation of clouds and cloudlets is not as significant in our case, though not negligible (see, e.g., \citealt{Kooij_2021}).

Also, we do not attempt to directly model the heating and ionization of gas from UV photons emitted by galaxies.
However, the effect of heating is
indirectly roughly taken into account at low temperature by the $T\sim 10^{4}\,\rm K$ cooling floor as mentioned in Section \ref{sec:method}.
At temperatures around $T\sim 10^{5}\,\rm K$, the inclusion of UV photons will reduce the cooling efficiency and as a result, the formation of cold clumps condensed from the cloud-halo mixing will be not as efficient as we see in models discussed in this paper.

\subsection{Milky Way HVCs}\label{sec:milky_way}

In this section, we apply our findings from the simulated clouds to explain and predict observable properties of Milky Way HVCs.

\subsubsection{Detectability on the RM sky}

So far searches for magnetized HVCs has been mainly focused on identifying coherent structure on the Faraday rotation measure (RM) grid that spatially overlaps with the distribution of observed HI clouds on the sky-plane (\citealt{McClure-Griffiths_2010}; \citealt{Hill_2013}; \citealt{Kaczmarek_2017}; \citealt{Betti_2019}; \citealt{Jung_2021}).
For a magnetized cloud to be able to leave a clear spatially resolved RM structure, it is favourable to have a coherent magnetic field over a large scale.
The more complex geometry of magnetic fields expected for clumpy clouds could significantly restrict the observational detectability of magnetized HVCs.

Instead, the imprint of magnetized HVCs could appear in the RM variance. The second-order structure function of RM is often used in earlier observational studies to parameterize scale-dependent turbulent magnetic field of the Galactic and extragalactic ISM (\citealt{Minter_1996}; \citealt{Haverkorn_2008}; \citealt{Stil_2011}; \citealt{Anderson_2015}; \citealt{Livingston_2021}; \citealt{Raycheva_2022}; \citealt{Seta_2023}).
Sightlines that intersect with magnetized HVCs can contribute to higher RM variance, as a result, change in the amplitude and the slope of the RM structure function. Angular scales of such excessive RM variance caused by magnetized HVCs would depend on various conditions, such as the distance to the clouds.
The RM grid density provided by currently available polarization data is not sufficient for such a statistical approach to the small-scale RM variation in HVCs. We expect upcoming polarization surveys, for example, the Polarisation Sky Survey of the Universe's Magnetism (ASKAP-POSSUM\footnote{https://askap.org/possum/}), will be capable of performing further investigations with significantly improved RM source density. 

\subsubsection{Decelerated structures}

In Section \ref{sec:fiducial_cloud} and Section \ref{sec:metal}, we find significant deceleration of clumps behind the main body of the clouds in our models.
Our models do not include the effect of Milky Way's gravitational potential, but as discussed earlier in Section \ref{sec:survivability}, \citet{Gronnow_2022} show that most of the clumps remain decelerated due to the magnetic tension even when the gravity is considered.
The presence of clumps with low relative velocity to the surrounding halo gas suggests that, observationally, HVCs are not limited to what we find in the high-observed-velocity range but likely to be associated with lower-observed-velocity components.
Indeed, \citet{Lockman_2008} and \citet{Hill_2013} identify several decelerated clumps in HI and H$\alpha$ emission respectively associated with the Smith cloud evident in the position-velocity space. 
We note that the models we use in this paper do not involve physics to differentiate various ISM phases (e.g., ionization) but rather treat the gas as a single-phase fluid with different temperatures. Therefore,
the investigation of the spatial distribution of neutral and ionized medium is beyond the scope of this paper.

\subsubsection{Growth of instabilities and small-scale structures}

Earlier in Fig. \ref{fig:all_model}, we have shown that clouds in our clumpy cloud models share roughly common morphological features. They all develop multiple leading heads attached to the main body and trailing filamentary tails regardless of the environment or physical conditions.
On top of that, the presence of the halo's magnetic field significantly suppresses the growth of small-scale structures and leads to the development of elongated filamentary clumps.
It is worth noting that it is the individual clumps that we find are elongated, not the entire cloud. The mass profile of the entire cloud is rather more extended along the direction of the cloud's motion in the HC model than in the MC model.

Observationally, such multi-scale analysis requires targeted high-quality observation data as the smallest detectable scale is restricted by the resolution and the signal-to-noise ratio.
Thus, the search for small-scale structures across the cloud-wind interface has mainly targeted HVC complexes with a large sky coverage.
The high-resolution Herschel image of the Draco nebula presented by \citet{Miville-Deschenes_2017} shows multiple heads (or ``fingers'' as they refer to them in the paper) linked to each other and pointing toward the Galactic plane. 
Similarly to what we see in our models, the separation between the dense heads is periodic as a result of the RT instability.
\citet{Barger_2020} identified an array of HI clumps hanging from the edges of Complex A. Their morphological and kinematical analysis explains that these structures are a snapshot of the active growth of hydrodynamic instabilities, both KH and RT. 
\citet{Marchal_2021} examined a small area of HVC Complex C and found multi-scale and multi-phase structures in the region. The identified clumps are predominantly elongated (the aspect ratio of $1.5-6$) and have preferred position angle orientations. Such highly elongated substructures favour the magnetized cloud in our models.

\subsubsection{The origin and fate of clouds}\label{sec:obs_metal}

Metallicity is often considered an indicator of the diverse origins of the HVCs.
Clouds formed from recycled materials of the Galactic fountain are expected to have close-to-solar metallicities (e.g., the Smith Cloud, intermediate-velocity clouds) and ones with extragalactic origin have the metallicity of primordial intergalactic medium if not enriched by external galaxies (e.g., the Magellanic system).
Models in this study show significant metallicity dependency on the clouds' morphology, the magnetic field strength, and the evolution of the clouds' cold gas mass.
Our results imply that HVCs with diverse origins would evolve differently under the same environment. 
Fountain-originated clouds are more efficient in radiative cooling than extragalactic-originated clouds which increases the chances of the clouds eventually reaching the Galactic gas disk and fueling the star formation activity.
A caveat here is that the metallicity of a cloud varies with time. Clouds and the surrounding halo gas can have different metallicity; usually, halo gas is expected to be more metal-poor than fountain-driven clouds. 
In \citet{Heitsch_2021}, the authors show steady replacement of cloud materials by the ambient halo gas when the mixing and the condensation are efficient, and therefore, the metallicity composing a cloud constantly changes with time.
This implies that the metal mixing at the cloud-halo boundary can alter the cooling efficiency over the course of the cloud evolution.

\section{Conclusion}\label{sec:summary}

Understanding the growth and disruption of cold clouds infalling from intergalactic space provides insight into whether they are capable of delivering cold gas to the galaxy and fueling star formation.
Many earlier numerical simulations that studied the survival of moving clouds in hot magnetized halo assume simple spherically symmetric uniform density distributions as an initial condition, whereas observed HVCs in similar environments clearly have non-uniform clumpy structures.
Motivated by this, we focus on performing simulations of initially clumpy clouds and examining whether the change of the cloud structure alters our understanding of the evolution of clouds in magnetized halos.

The total cold gas mass in the uniform-density clouds grows from the beginning as expected from earlier models with similar settings (\citealt{Gronnow_2018}; \citealt{Kooij_2021}), while the clumpy clouds lose cold gas early on instead. Still, later on, even clumpy clouds experience a significant growth of mass. Hence, the finding of the earlier works that infalling HVCs might be able to grow and feed the Galaxy still holds to clumpy clouds.
Our results are consistent with what is discussed in other numerical studies modelling the ISM under hot galactic feedback outflows (\citealt{Gronke_2018}; \citealt{Banda-Barragan_2021}) or clouds launched from the ISM into the halo (i.e., the galactic fountain; \citealt{Armillotta_2016}). In both cases, the cloud's evolution is simulated under similar settings to our models, but they are in a regime where the turbulent structure of clouds is better constrained (i.e., the ISM) compared to clouds infalling from the intergalactic space.

In the fiducial model, some massive clumps to some degree have similar characteristics with uniform density clouds. They have a head-tail morphology and magnetic field lines draped around each of them.
In comparison, these characteristics are lost among clumps trailing behind the cloud. 
The configuration of magnetic fields is dominated by a large-scale pattern that drapes around the entire cloud system. 
Clumps in this downstream region face magnetic fields that are almost parallel to the direction of the flow, therefore, the magnetic field draping around these clumps is inefficient. 
These clumps are mostly condensed from the halo gas and significantly decelerated from the high relative velocity of the main cloud.

By comparing the structural properties of clouds evolved under different conditions, we study the significance of ram pressure stripping, growth of hydrodynamic instabilities, and radiative cooling.
Specifically, we compare models (i) with and without halo magnetic fields, (ii) with varying metallicity and cooling efficiency, and (iii) with different slopes of the initial density power spectrum.
\begin{enumerate}
\item With magnetic fields present, there is an indication of suppressed hydrodynamic instabilities at the cloud-halo interface along the direction of the cloud's motion.
The magnetized cloud (MC model) has clumps that in general have large effective sizes along the direction of motion.
In comparison, its non-magnetized counterpart (HC model) is composed of a larger number of smaller clumps. This is because clumps break into smaller pieces via the active growth of hydrodynamic instabilities rather than being elongated along the streaming direction.
Likewise, the 2D spatial power spectrum (SPS) of magnetized clouds has less power at small length scales compared to that of non-magnetized clouds throughout the time span covered in this work.

\item Radiative cooling efficiency and the metallicity of the cloud have two apparent effects on the cloud structure.
When cooling is efficient (metallicity is higher), first, a greater number of dense clumps condense from the halo gas in the wake behind a cloud. These clumps are substantially decelerated with respect to the surrounding halo gas compared to the initial high-velocity motion of the cloud. 
Second, the distributions of density and magnetic field lines are more compact.
This is because the intermediate density inter-clump material, which would have been easily pushed away by the ram pressure when cooling was not efficient, cools down to lower temperatures (i.e., higher density) and acts like a bond between the clumps. We demonstrate that this effect leads to a change in the overall pressure gradient and the flow pattern around a clumpy cloud.

\item The slope of the initial density spatial power spectrum defines the relative strength between small- and large-scale structures in a clumpy cloud.
A cloud with a shallower power spectrum slope (MC-p2.7) has a wider density PDF, i.e., higher density contrast between clumps and inter-clump gas compared to its steeper slope counterpart (MC-p3.5).
Early in the simulation, the total cold gas mass in the cloud with the shallower slope decreases faster because the ram pressure can easily push the low-density inter-clump gas.
On the other hand in the later stage of cloud evolution, such stripped intermediate-density gas mixes with the halo gas and cools down to form dense filamentary tails behind the cloud resulting in significant growth of the cold gas mass (e.g., \citealt{Armillotta_2016}; \citealt{Gronke_2018, Gronke_2020}; \citealt{Li_2020}; \citealt{Kanjilal_2021}; \citealt{Banda-Barragan_2021}; \citealt{Heitsch_2021}; \citealt{Gronnow_2022}).
In comparison, the cloud with a steeper power spectrum shows less significant mass loss at the beginning followed by a more delayed upturn of mass. 
\end{enumerate}

To summarise, our results indicate that the progression of processes that shape the evolution of high-velocity clouds (ram pressure stripping, hydrodynamic instabilities, and radiative cooling) is closely linked to the presence of magnetic fields as well as clumpy substructures within the clouds.

\section*{Data availability}
The data directly related to this article will be shared on reasonable request to the corresponding author.

\section*{Acknowledgements}

We thank the referee for their helpful report which improved the paper. We acknowledge high-performance computing resources provided by the Australian National Computational Infrastructure (project ix81) in the framework of the ANU Merit Allocation Scheme. N. M. Mc-G acknowledges funding from the Australian Research Council in the form of DP190101571 and FL210100039.

Our analysis was performed using the Python programming language (Python Software Foundation, https://www.python.org). The following packages were used throughout the analysis: numpy (\citealt{Harris_2020}),  SciPy (\citealt{Virtanen_2020}), and matplotlib (\citealt{Hunter_2007}). This research also made use of publicly available packages and software including
VisIt \citep{VisIt}, yt project \citep{yt_project}, {\sc powerbox} \citep{Murray_2018}, 
{\sc Turbustat} \citep{Koch_2019}, 
{\sc CUPID} \citep{Berry_2007}.

S.L.J. and N. M. Mc-G acknowledge the Ngunnawal and Ngambri people as the traditional owners and ongoing custodians of the land on which the Research School of Astronomy \& Astrophysics is sited at Mt Stromlo.

\section*{Links to movies}
Movies of the simulations are available here.
MC model (\href{https://youtube.com/watch?v=kmizSlb7pVE?feature=share}{Link}); 
HC model (\href{https://youtube.com/watch?v=ty28gNGqj7Y}{Link});
MU model (\href{https://youtube.com/watch?v=xGewhkfrUZc?feature=share}{Link});
HU model (\href{https://youtube.com/watch?v=9-a3zsjHPqc}{Link});
MC-Z0.03 model (\href{https://youtube.com/watch?v=FXxUHBmzCY0?feature=share}{Link});
MC-NoCool model (\href{https://youtube.com/watch?v=pqIGOWbzPao?feature=share}{Link});
MC-p2.7 model (\href{https://youtube.com/watch?v=6xcvDoexMWk?feature=share}{Link});
MC-p3.5 model (\href{https://youtube.com/watch?v=Hgx2ARaH5zg?feature=share}{Link})



\bibliographystyle{mnras}
\bibliography{mybib} 

\bsp	
\label{lastpage}
\end{document}